\PassOptionsToPackage{unicode}{hyperref}
\PassOptionsToPackage{hyphens}{url}
\PassOptionsToPackage{dvipsnames,svgnames,x11names}{xcolor}
\documentclass[
  number,
  preprint]{elsarticle}
\usepackage{xcolor}
\usepackage{amsmath,amssymb}
\usepackage{iftex}
\ifPDFTeX
  \usepackage[T1]{fontenc}
  \usepackage[utf8]{inputenc}
  \usepackage{textcomp} 
\else 
  \usepackage{unicode-math} 
  \defaultfontfeatures{Scale=MatchLowercase}
  \defaultfontfeatures[\rmfamily]{Ligatures=TeX,Scale=1}
\fi
\usepackage{lmodern}
\ifPDFTeX\else
\fi
\IfFileExists{upquote.sty}{\usepackage{upquote}}{}
\IfFileExists{microtype.sty}{
  \usepackage[]{microtype}
  \UseMicrotypeSet[protrusion]{basicmath} 
}{}
\makeatletter
\@ifundefined{KOMAClassName}{
  \IfFileExists{parskip.sty}{%
    \usepackage{parskip}
  }{
    \setlength{\parindent}{0pt}
    \setlength{\parskip}{6pt plus 2pt minus 1pt}}
}{
  \KOMAoptions{parskip=half}}
\makeatother
\makeatletter
\ifx\paragraph\undefined\else
  \let\oldparagraph\paragraph
  \renewcommand{\paragraph}{
    \@ifstar
      \xxxParagraphStar
      \xxxParagraphNoStar
  }
  \newcommand{\xxxParagraphStar}[1]{\oldparagraph*{#1}\mbox{}}
  \newcommand{\xxxParagraphNoStar}[1]{\oldparagraph{#1}\mbox{}}
\fi
\ifx\subparagraph\undefined\else
  \let\oldsubparagraph\subparagraph
  \renewcommand{\subparagraph}{
    \@ifstar
      \xxxSubParagraphStar
      \xxxSubParagraphNoStar
  }
  \newcommand{\xxxSubParagraphStar}[1]{\oldsubparagraph*{#1}\mbox{}}
  \newcommand{\xxxSubParagraphNoStar}[1]{\oldsubparagraph{#1}\mbox{}}
\fi
\makeatother

\usepackage{longtable,booktabs,array}
\usepackage{calc} 
\usepackage{etoolbox}
\makeatletter
\patchcmd\longtable{\par}{\if@noskipsec\mbox{}\fi\par}{}{}
\makeatother
\IfFileExists{footnotehyper.sty}{\usepackage{footnotehyper}}{\usepackage{footnote}}
\makesavenoteenv{longtable}
\usepackage{graphicx}
\makeatletter
\newsavebox\pandoc@box
\newcommand*\pandocbounded[1]{
  \sbox\pandoc@box{#1}%
  \Gscale@div\@tempa{\textheight}{\dimexpr\ht\pandoc@box+\dp\pandoc@box\relax}%
  \Gscale@div\@tempb{\linewidth}{\wd\pandoc@box}%
  \ifdim\@tempb\p@<\@tempa\p@\let\@tempa\@tempb\fi
  \ifdim\@tempa\p@<\p@\scalebox{\@tempa}{\usebox\pandoc@box}%
  \else\usebox{\pandoc@box}%
  \fi%
}
\def\fps@figure{htbp}
\makeatother

\NewDocumentCommand\citeproctext{}{}

\makeatletter
 \let\@cite@ofmt\@firstofone
 \def\@biblabel#1{}
 \def\@cite#1#2{{#1\if@tempswa , #2\fi}}
\makeatother
\newlength{\cslhangindent}
\newlength{\csllabelwidth}
\newenvironment{CSLReferences}[2] 
 {\begin{list}{}{%
  \setlength{\itemindent}{0pt}
  \setlength{\leftmargin}{0pt}
  \setlength{\parsep}{0pt}
  \ifodd #1
   \setlength{\leftmargin}{\cslhangindent}
   \setlength{\itemindent}{-1\cslhangindent}
  \fi
  \setlength{\itemsep}{#2\baselineskip}}}
 {\end{list}}
\usepackage{calc}

\usepackage{booktabs}
\usepackage{caption}
\usepackage{longtable}
\usepackage{colortbl}
\usepackage{array}
\usepackage{anyfontsize}
\usepackage{multirow}
\usepackage{booktabs} \usepackage{longtable} \usepackage{placeins} \usepackage{hyperref} \hypersetup{ colorlinks=true, linkcolor=blue, citecolor=blue, urlcolor=blue } \newpageafter{abstract}
\makeatletter
\@ifpackageloaded{caption}{}{\usepackage{caption}}
\AtBeginDocument{%
\ifdefined\contentsname
  \renewcommand*\contentsname{Table of contents}
\else
  \newcommand\contentsname{Table of contents}
\fi
\ifdefined\listfigurename
  \renewcommand*\listfigurename{List of Figures}
\else
  \newcommand\listfigurename{List of Figures}
\fi
\ifdefined\listtablename
  \renewcommand*\listtablename{List of Tables}
\else
  \newcommand\listtablename{List of Tables}
\fi
\ifdefined\figurename
  \renewcommand*\figurename{Figure}
\else
  \newcommand\figurename{Figure}
\fi
\ifdefined\tablename
  \renewcommand*\tablename{Table}
\else
  \newcommand\tablename{Table}
\fi
}
\@ifpackageloaded{float}{}{\usepackage{float}}
\floatstyle{ruled}
\@ifundefined{c@chapter}{\newfloat{codelisting}{h}{lop}}{\newfloat{codelisting}{h}{lop}[chapter]}
\floatname{codelisting}{Listing}

\makeatother
\makeatletter
\@ifpackageloaded{caption}{}{\usepackage{caption}}
\@ifpackageloaded{subcaption}{}{\usepackage{subcaption}}
\makeatother
\journal{Preprint}
\usepackage{bookmark}
\IfFileExists{xurl.sty}{\usepackage{xurl}}{} 
\hypersetup{
  pdftitle={Little Impact of ChatGPT on High School Test Scores},
  pdfauthor={Nick Huntington-Klein},
  pdfkeywords={AI, ChatGPT, Test Scores, Measurement, Observational},
  colorlinks=true,
  linkcolor={blue},
  filecolor={Maroon},
  citecolor={Blue},
  urlcolor={Blue},
  pdfcreator={LaTeX via pandoc}}

\begin{document}

\begin{frontmatter}
\title{Little Impact of ChatGPT on High School Test Scores}
\author[1]{Nick Huntington-Klein%
\fnref{fn1}}
 \ead{nhuntington-klein@seattleu.edu} 

\affiliation[1]{organization={Seattle University, Department of
Economics},city={Seattle, WA},postcodesep={}}

\cortext[cor1]{Corresponding author}
\fntext[fn1]{Associate Professor of Economics, Seattle University.
Research assistance for this project comes from Seamus Kelly and Loan
Tran. The author has no relevant financial support related to this
research to report.}
        
\begin{abstract}
In educational settings, AI can be used as a learning aid, but can also
be used to avoid schoolwork, passing classes while learning little. Most
existing evidence on AI's educational impact comes from trials of
supervised instructor-encouraged AI usage, rather than on unsupervised
usage which is of the most concern. This paper uses the dropoff in
ChatGPT activity during non-school summer months to identify areas with
heavy educational AI use, and then estimates the effect of that usage on
aggregate United States high school test scores. Areas with higher
educational AI use were not significantly different from other areas,
with an insignificant effect estimated to be close to zero, and in the
preferred specification bounded below roughly 0.04 standard deviations
of test score performance in absolute terms.
\end{abstract}

\end{frontmatter}

\section{Introduction}\label{introduction}

Possibly the most pressing question in education research as of the
mid-2020s is what the impact of large language models (LLMs) will be on
learning. There are reasons to expect both positive and negative impact.
LLMs open up the possibility of personalized instruction on a scale no
school system could previously staff. LLMs also make it trivially easy
for students to turn in high-quality work without learning anything
themselves.

Which of these effects dominates is an empirical question. These
mechanisms also point to different kinds of artificial intelligence (AI)
usage. Most quantitative work on the learning impacts of AI is
experimental and evaluates a specific intentional AI application or
tool, deployed by an instructor, usually in a single college course
(Deng et al. 2025). That literature speaks to whether intentional
in-class ChatGPT usage can improve outcomes. It does not speak to
whether student-directed AI use, as it actually occurs, is changing how
much students learn. These are different treatments.

This paper focuses on aggregate student AI usage, including unsupervised
and independent AI usage, at the high school level in the United States,
where roughly 80\% of students reported using generative AI tools for
schoolwork as of early 2025 (Adair et al. 2025) and where causal
evidence is limited. The empirical strategy exploits a feature of
aggregate ChatGPT usage: activity falls sharply during the summer. Areas
where activity falls further in the summer are areas where a larger
share of ChatGPT use is tied to school.

Regional variation in the ``summer dip'' in ChatGPT usage and activity
is then related to changes in high school test scores, combining a large
number of different federal, state, and local tests. Estimated effects
are generally small and insignificant. In the main model, one standard
deviation increase in educational ChatGPT exposure changes high school
test scores by -0.0096 (0.0098) standard deviations in the baseline
model. This general finding is robust to many different modeling
decisions, and variants where the estimated effects are significant are
not themselves robust. Under reasonable assumptions about measurement
error in the treatment variable, the effect can be bounded to be no
worse than -0.04 standard deviations. This null finding is not
consistent with the idea that the availability of AI tools has
meaningfully harmed aggregate student performance, although it cannot
rule out small negative effects.

This estimate comes in a setting where ideal data is not available to
researchers. As such, a primary task of the paper is to validate that
the measurement does reflect usage of ChatGPT by high school students,
and that the identifying assumptions are reasonable. The paper contains
a long list of validation exercises which generally support the use of
the summer-dip measurement, but measurement remains imperfect. However,
this paper addresses an important gap in the literature: there is little
evidence looking at the impact of LLMs as they are actually used by high
school students. Given that many of the potential concerns about AI
usage concern topics like cheating, which would not be encouraged by
education-specific AI tools, the task of using what data is available to
evaluate the impact of unsupervised AI usage is worthwhile.

\section{Related literature}\label{sec-lit}

\subsection{Educational impacts of purpose-built AI
tools}\label{educational-impacts-of-purpose-built-ai-tools}

A great deal has been written on AI in education very quickly. A search
of OpenAlex suggests roughly 15,000 papers between 2023 and
mid-2026.\footnote{Search performed on May 5, 2026 for papers tagged in
  OpenAlex's ``Education'' concept with ``ChatGPT'', ``large language
  model'', or ``generative AI'' in title or abstract. More than 99\%
  postdate the release of ChatGPT.} Most of the quantitative work is
experimental. (Deng et al. 2025) meta-analyze the 2022 to 2024
experimental evidence, mostly on college students, and find largely
positive effects of AI instructional tools on performance, while noting
that in many cases these studies use small samples.

Unsurprisingly for such a large and fast-moving topic of study, on a
treatment like ``AI'' that can mean a wide range of different things,
results vary significantly across studies. (Kestin et al. 2025) find
positive effects of a custom AI tutor relative to in-class active
learning in college physics, and (Chan, Lo, and Wong 2024) and (Zhang
2025) find similar results for college writing. (De Simone et al. 2025)
evaluate a six-week after-school program in Nigeria using an LLM-based
tutor and find a large effect of 0.23 standard deviations on
English-language learning performance among secondary-school students.
Other studies find that the introduction of AI tools harms performance,
or offers temporary gains that eventually reduce performance once access
to the tool is removed (Barcaui 2025; Shi et al. 2025).

One repeated theme in these experiments is that additional guidance on
AI usage can make tools more effective at conferring actual learning.
(Bastani et al. 2024) find strong performance in high school mathematics
while students have AI access and negative effects once it is withdrawn
unless AI integrations had appropriate guardrails. (Gallegos 2026)
deploys a course-specific AI assistant in undergraduate econometrics in
a randomized encouragement study, finding that adoption alone does not
improve exam performance, but that additional guidance on tool usage
improved exam performance.

\subsection{Observational evidence on educational AI
impacts}\label{observational-evidence-on-educational-ai-impacts}

Causal observational work, which allows researchers to evaluate the
impacts of AI as it is used and outside of a controlled environment, is
thinner.

(Hausman, Rigbi, and Weisburd 2025) use cross-sectional variation in how
AI-compatible a course is at an Israeli university. They find that AI
availability raises grades, especially for lower-performing students,
and compresses the grade distribution, with evidence of gains in
AI-specific human capital but possible losses in traditional human
capital. Similarly, (Chirikov 2026) examines grades at a large research
university and compares courses where assessed tasks are more or less
exposed to AI. The share of A grades rose by 13 percentage points in
high-exposure courses after ChatGPT's release, about 30\% relative to
the 2022 baseline. The increase was concentrated in courses where a
heavier share of the grade came from homework, suggesting that the
improvement was not related to actual learning that would improve exam
scores.

Large-scale observational work for K-12 students, rather than college
students, is less common. (Strömberg, Lei, and Wu 2026) observe 26,811
Chinese students in grades 7 through 12 over 30 months, using a
staggered-adoption difference-in-differences design to evaluate the
impact of AI adoption. They find that AI adoption raises homework scores
by 18\% and cuts completion time by 30\%, but lowers monthly examination
scores by 20\% within six months. Entrance examination scores fall by 18
to 24\%, with this impact developing slowly and only fully present after
two years. Deeper analysis on AI usage finds that the negative effects
were concentrated on students whose usage and short homework times
indicated they were outsourcing their work; however, this is estimated
to be 80\% of AI users.

(Yu et al. 2024) study a large number of academic writing submissions
before and after ChatGPT's release. They find quality gains concentrated
at the bottom of the distribution, compressing the quality distribution.

Some studies are able to partially address selection into the use of AI
tools. (Hanshaw, Wicker, and Denlinger 2025) uses propensity score
matching to address some forms of selection bias, finding positive
propensity-adjusted relationships between the use of an AI tutoring tool
and learning, while (Dunn 2025) accounts for some selection pressures
using a before/after design, finding mixed impacts on performance in a
college-level statistics class. (McNichols, Ikram, and Lan 2025) have
access to student AI usage logs in a college-level class; students who
used the AI for assistance in understanding saw improved understanding,
while those who used the AI to avoid work performed more poorly. (Wecks
et al. 2024) use AI detection tools to measure student-level variation
in AI usage, with course-of-study-level fixed effects and controls,
finding strong negative relationships between AI usage and exam scores
in a college-level accounting class.

\section{Measuring educational ChatGPT use}\label{sec-measure}

This project brings together several different data sets at several
different levels of granularity. Most of the consequential research
decisions made in this study come at the level of data preparation and
measurement, rather than in analytical design or estimation. This
section constructs measurements of the treatment variable, aggregate
educational AI usage, and then subjects it to three tests: whether it
tracks school calendars, whether it is driven by college users, and
whether it is reliable enough to support cross-sectional comparisons.

\subsection{Data on ChatGPT activity}\label{data-on-chatgpt-activity}

Two sources measure ChatGPT activity. Similarweb is a web traffic data
company and tracks web traffic to chatgpt.com and openai.com. These are
directly interpretable absolute counts of the number of visits to these
sites, but this data is only available in aggregate for the country, or
on the state level for a limited set of 15 states. The second source of
ChatGPT activity is Google Trends, which reports an index of search
interest for the term ``chatgpt'', weekly, for all states and 210
metropolitan areas. Google Trends is the primary source because of its
geographic coverage. Because of Similarweb's limited coverage, its
primary use here is to show that Google Trends closely tracks actual web
activity. See Section~\ref{sec-similarweb} for comparisons of the two
sources.\footnote{Details on how all the different test score and web
  activity sources are combined together are in
  Section~\ref{sec-merging}.}

The Google Trends index requires care in interpretation. Rather than
reporting absolute search volumes, Google Trends produces an index of
search activity relative to the total number of Google searches in that
region and time (Stephens-Davidowitz and Varian 2014). The purpose of
this study is to examine relative within-region search activity, and so
the use of a relative indicator of search activity is not inherently
problematic. Further, the calculation of summer relative to school-year
activity is performed separately by year, so the possibility that the
index might be affected by a change in the overall number of Google
searches over time is minimized.

There are two particularly important mechanical issues with Google
Trends. First, Google Trends censors low-volume cells to zero. This does
not occur in the state-level data but occurs in 2,038 of 32,970
metropolitan-week observations in 2023 through 2025. The impact of
zero-censored observations is reduced because the index is anchored in
2021, when ChatGPT activity was truly near 0, pushing later values in
the analysis window away from the lower censoring threshold. The
remaining 0s in the 2023-2025 period are true low values and so are
included as their censored 0 value; Table~\ref{tbl-calendar} shows that
results are robust to dropping the 0s.

\subsection{The summer dip measure}\label{the-summer-dip-measure}

Educational AI usage is measured as the ratio of average activity while
school is in session to average activity in the summer. There is
considerable variation across schools and regions in terms of when
exactly summer begins and ends and when schools go on non-summer breaks,
but since the goal is a reliable annual measure comparing summer usage
vs.~in-session usage, this measure does not attempt to use exact session
start and end dates, and instead uses conservative time windows shared
across schools to ensure summer and in-session times are being compared.
In-session is defined as January through April and 16 September through
18 December, and summer is defined as 16 June through 15 August. The
remaining weeks are dropped as periods when schools are mixed, or in
late December's case when schools are on holiday.

One key caveat to the summer dip measure is that it cannot separate K-12
academic use from any other use following the same seasonal pattern,
most obviously college students. This is explored below.

\subsection{Do school calendars drive the measure?}\label{sec-calendar}

While the summer dip measure does not rely on sharp term start/end
dates, as a gut check the Google Trends data should sharply follow term
start/end dates where they can be verified. This section examines the
quality of the summer dip measure by checking whether it appears to line
up with K-12 activity, using start/end dates to verify that
relationship.

American K-12 calendars show enough variation that this can be tested.
Schools open in the first week of August in Alabama, Arizona, Georgia
and Tennessee, and after Labor Day in Minnesota, New Jersey, New York
and Wisconsin. They close in mid-May in Arizona and Missouri, and in
late June in Maine, New York and Rhode Island. Standard college semester
and quarter calendars are more compressed and do not follow that
geography, so a measure that moves with local K-12 timing is difficult
to attribute to college students.

\subsubsection{Summer-drop validity against more precise calendar
data}\label{sec-caldata}

More-precise calendar data on school start and end dates comes from a
published start-and-end-date chart (Clell Wade Coaches Directory 2026),
which sorts each state into half-month windows based on 2024-2025 data,
shown in Table~\ref{tbl-calgroups}. These data are consistent with
regional opening date distributions from (Pew Research Center 2023).
Importantly, these are state level measures of when opening dates tend
to be for the majority of schools in a state; within-state variation in
opening and closing dates is not captured.

\begin{figure}

\centering{

\pandocbounded{\includegraphics[keepaspectratio]{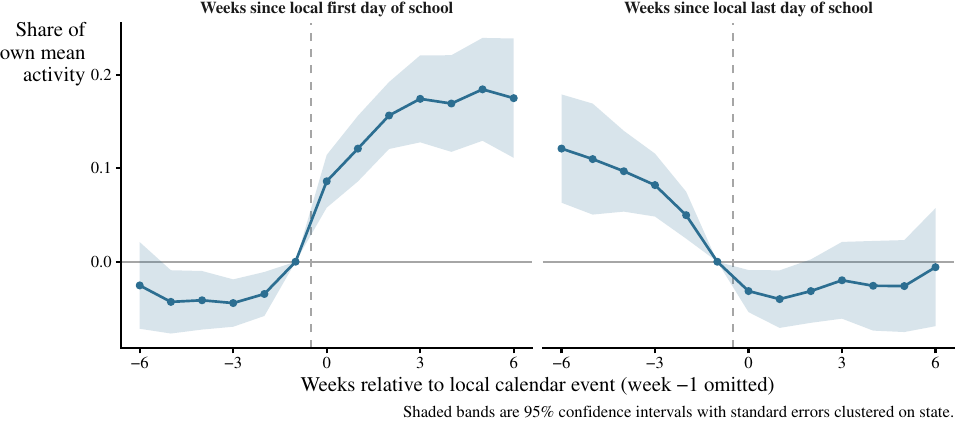}}

}

\caption{\label{fig-caleventstudy}ChatGPT activity breaks at each area's
own calendar dates}

\end{figure}

Figure~\ref{fig-caleventstudy} shows ChatGPT search activity lined up
against local school start and end dates. Activity picks up specifically
at the week that school starts, in the left panel, after several weeks
of flat activity. In the right panel, there is a slow decline in
activity as schools near the end of their term, followed by a brief flat
period before some growth in the summer. An important distinction
between the end-of-school and start-of-school analyses is that ChatGPT
activity is generally trending upwards over time, which can explain the
during-summer upward trend after school lets out.

Additional results in the appendix reinforce the finding that activity
is specifically moving around K-12 start and end dates.
Figure~\ref{fig-profile} and Figure~\ref{fig-profile-end} show the
pattern in raw ChatGPT Google Trends activity. Figure~\ref{fig-profile}
normalizes each state-year to its own July mean while
Figure~\ref{fig-profile-end} normalizes to the April mean. The
start-date and end-date groups largely separate in the expected
direction: the earlier the school year starts/ends, the earlier ChatGPT
activity picks back up/drops off.

I statistically test the alignment between the Google Trends index for
``ChatGPT'' and the K-12 school calendar using the specification:

\begin{equation}\phantomsection\label{eq-calendar}{
Trends_{iwy} = \alpha_{iy}+\delta_{wy}+ \beta InSession_{iw} + \varepsilon_{iwy}
}\end{equation}

where \(Trends_{iwy}\) is the Google Trends index for ``ChatGPT'' in
area \(i\) (either a state or metropolitan area, depending on analysis)
in calendar week-of-year \(w\) of year \(y\), and \(InSession_{iw}\) is
a binary indicator for whether \(w\) is an in-session week for area
\(i\), based on calendars for most schools in the state that \(i\) is
in. The specification includes area-by-year fixed effects
\(\alpha_{iy}\) and week-by-year fixed effects \(\delta_{wy}\) which
account for regional and seasonal variation in ChatGPT usage that is
unrelated to K-12 use, as well as the general increasing popularity of
ChatGPT over time. \(\beta\) compares areas where the schools in that
area's state are in-session against areas that are predicted to be
out-of-session, in the same week of the year. Standard errors are
clustered at the state level.

\begin{table}

\caption{\label{tbl-calendar}Effect of local schools being in session on
ChatGPT search interest}

\centering{

\fontsize{7.0pt}{8.0pt}\selectfont
\begin{tabular*}{\linewidth}{@{\extracolsep{\fill}}>{\raggedright\arraybackslash}p{\dimexpr 0.40\linewidth -2\tabcolsep-1.5\arrayrulewidth}rr>{\raggedright\arraybackslash}p{\dimexpr 0.26\linewidth -2\tabcolsep-1.5\arrayrulewidth}}
\toprule
Specification & Estimate & Obs & Note \\ 
\midrule\addlinespace[2.5pt]
Fall, states & 0.127 (0.017)*** & 1,785 &  \\ 
Fall, states, dropping transition weeks & 0.176 (0.025)*** & 1,607 & Removes partial-session opening weeks \\ 
Fall, metropolitan areas & 0.151 (0.043)*** & 7,297 &  \\ 
Fall, metropolitan areas, dropping transition weeks & 0.207 (0.046)*** & 6,555 &  \\ 
Fall, single-state metropolitan areas only & 0.147 (0.047)*** & 6,562 & Drops metros crossing multiple states \\ 
Fall, dropping censored zeros & 0.106 (0.015)*** & 6,720 & metropolitan areas \\ 
Spring, states & 0.080 (0.015)*** & 1,581 & end-of-year timing \\ 
Spring, states, dropping transition weeks & 0.095 (0.020)*** & 1,377 & Removes partial-session closing weeks \\ 
Spring, metropolitan areas & 0.105 (0.045)** & 6,463 &  \\ 
Spring, metropolitan areas, dropping transition weeks & 0.060 (0.045) & 5,629 &  \\ 
Spring, single-state metropolitan areas only & 0.104 (0.048)** & 5,812 & Drops metros crossing multiple states \\ 
June only, states & 0.079 (0.019)*** & 459 & Most college terms ended \\ 
June only, states, dropping transition weeks & 0.120 (0.030)*** & 359 & Removes partial-session closing weeks \\ 
June only, metropolitan areas & 0.061 (0.031)* & 1,877 &  \\ 
June only, metropolitan areas, dropping transition weeks & 0.052 (0.067) & 1,519 &  \\ 
June only, single-state metropolitan areas only & 0.060 (0.034)* & 1,688 & Drops metros crossing multiple states \\ 
Early August (8 Jul to 17 Aug), states & 0.048 (0.014)*** & 561 & No US college in session \\ 
Early August (8 Jul to 17 Aug), metropolitan areas & -0.022 (0.075) & 2,293 & Short window \\ 
\bottomrule
\end{tabular*}
\begin{minipage}{\linewidth}
Notes: Coefficient shown is on being in-session. Each area-year series is divided by its own mean before estimation, because Google rescales every series to its own maximum and index points are therefore not comparable across areas. A coefficient of 0.10 means activity is 10 percent higher than that area typically records in that year while schools are in session. All specifications include area-by-year and week-of-year by year fixed effects, using 2023-2024 data.Standard errors clustered on state. Significance: * p<0.10, ** p<0.05, *** p<0.01.\\
\end{minipage}

}

\end{table}%

Table~\ref{tbl-calendar} reports the results. Using state-level data,
activity is 0.127 (0.017)*** higher while local schools are in
session,\footnote{Throughout the paper, significance stars for in-line
  text match the convention used in the tables: * \(p < .1\), **
  \(p < .05\), *** \(p < .01\).} or about 13\% of an area's typical
activity in that window. Dropping the partial transition week, for which
the impact on ChatGPT activity is uncertain, raises the estimate to
0.176 (0.025)***. The result holds at the metropolitan-area level, at
0.151 (0.043)***. The result is robust to dropping any of the calendar
groups (Table~\ref{tbl-loo}).

We do see weaker impacts of being in-session in the spring, as schools
are letting out, as opposed to in the fall when school is starting.
Based on Figure~\ref{fig-caleventstudy} this may be due to a slow
wind-down in ChatGPT activity near the end of the year. For this reason
I also look at June-only data, which is more poorly-balanced (most
schools are already out of session) but has a cleaner comparison. This
sample shows a sharper distinction, and dropping transition weeks
increases the coefficient size at the state level, as in fall and as
expected. Significance is lost only at the metropolitan level, where the
estimate is 0.052 (0.067); the state-level estimate remains significant
at 0.120 (0.030)***.

These results are consistent with Google Trends activity picking up
meaningful signal in K-12 usage, since usage does not just track overall
seasonality or regional differences, but specific differences in local
school calendars.

\subsubsection{Do college students explain the summer
dip?}\label{sec-college}

An obvious potential explanation of the summer dip is that it is driven
by college students, who follow a similar seasonal pattern to K-12
students. Section~\ref{sec-caldata} is evidence against this, since
cross-state variation in college calendars is not aligned with K-12
calendars. However, college students almost certainly contribute to the
measured summer dip overall. This could threaten identification if the
geographic variation in college student use is correlated with
geographic variation in K-12 usage, or could threaten measurement
quality if college student usage adds noise to the treatment variable.

I scale Google Trends indices by own-area means and interact the
in-session indicator with an area's college student share, scaled so
that one unit is the move from the 10th to the 90th percentile across
areas. The interaction coefficient is then 0.122 (0.040)*** in the fall
window against a main effect of 0.104 (0.047)**. The in-session response
rises from about 10 percent of an area's typical activity at the 10th
percentile of college share to about 23 percent at the 90th, a ratio of
roughly 2.2 to one. The same interaction is 0.045 (0.035) in the spring
window, which is not statistically distinguishable from zero, and -0.036
(0.027) in June, where the main effect is 0.080 (0.033)**.

Taking these results together, they establish that college students
cannot fully explain the summer dip, but that they do matter, especially
when comparing summer activity to fall activity. Linearly predicting
summer-dip measures with college student share shows that college share
accounts for 10.0 percent of the cross-area variance in the dip. To
address both potential confounding and the contribution of measurement
noise, the main results table includes an additional robustness check
that residualizes the exposure measure on college share. The
residualized version of the treatment variable does not qualitatively
change the results.

\subsection{Consistency of the measure}\label{sec-reliability}

Another way to both evaluate the amount of signal in the measure and to
evaluate the degree of noise is to evaluate its stability from year to
year. Notably, it is realistic that the values are truly different in
two years, so reliability below 1 does not indicate by itself a
measurement failure. The correlation between each metro's 2023 summer
dip and its 2024 summer dip is 0.19 (Pearson), or 0.65 (rank Spearman).
The deviation between the two suggests high year-to-year reliability but
with high noise outliers, driven by smaller metros, some of which report
summer dip ratios above seven. Winsorizing each year at the 5th and 95th
percentiles, which I pursue for the rest of the paper to avoid the
influence of these unrealistic outlier measurements, raises the overall
correlation to 0.68.

Stability is also related to an area's mean index level in 2023 and
2024, shown in Table~\ref{tbl-reliability}, ranging from -0.02 in the
lowest quartile to 0.91 in the highest. Because the index is scaled
between 0 and 100, with nearly every metro's peak value occurring in
2025, a high average 2023-2024 value indicates a metro where ChatGPT was
adopted early, with relatively more growth 2022-2023 than 2023-2025
relative to other metros. Being an early adopter is related to the share
of college students as well (a correlation of 0.41 between the index
level and college student share). Higher index levels are related to
higher reliability for two reasons. One is that weekly noise is lower
for higher-average metropolitan areas.\footnote{I repeatedly (40 times)
  randomly split each metro/year into two random halves and estimate the
  summer dip twice. Reliability rises monotonically across quartiles of
  the index level, from 0.50 to 0.83, against 0.61 overall.} The second
is that, because relatively more of the growth occurred before 2023,
there is meaningfully less year-to-year movement in the actual dip in
higher-average areas, which is indistinguishable from increased
2023-to-2024 reliability. As such, I add robustness tests throughout the
paper limiting the analysis to metropolitan areas with above-median
average index values in 2023-2024, where reliability is higher, noting
that this combines a precision limitation with a metro's status as an
early adopter.

\begin{table}

\caption{\label{tbl-reliability}Reliability of the summer dip measure
across metropolitan areas}

\centering{

\fontsize{7.0pt}{8.0pt}\selectfont
\begin{tabular*}{\linewidth}{@{\extracolsep{\fill}}lr}
\toprule
Row & Value \\ 
\midrule\addlinespace[2.5pt]
\multicolumn{2}{l}{A. Year-to-year correlation, 2023 vs 2024} \\[2.5pt] 
\midrule\addlinespace[2.5pt]
Winsorized at 0.0\% & 0.185 \\ 
Winsorized at 1.0\% & 0.419 \\ 
Winsorized at 2.5\% & 0.582 \\ 
Winsorized at 5.0\% & 0.678 \\ 
Winsorized at 10.0\% & 0.725 \\ 
Rank (Spearman) correlation & 0.648 \\ 
\midrule\addlinespace[2.5pt]
\multicolumn{2}{l}{B. Year-to-year correlation by measurement precision} \\[2.5pt] 
\midrule\addlinespace[2.5pt]
Precision quartile 1 (mean index 14.3, n = 41) & -0.017 \\ 
Precision quartile 2 (mean index 18.5, n = 52) & 0.756 \\ 
Precision quartile 3 (mean index 21.5, n = 52) & 0.896 \\ 
Precision quartile 4 (mean index 26.6, n = 52) & 0.911 \\ 
\midrule\addlinespace[2.5pt]
\multicolumn{2}{l}{C. Reliability of the two-year average exposure measure} \\[2.5pt] 
\midrule\addlinespace[2.5pt]
All metros & 0.798 \\ 
Higher-precision metros (top half) & 0.925 \\ 
\bottomrule
\end{tabular*}
\begin{minipage}{\linewidth}
Notes: Panel A correlates each area's 2023 summer dip with its 2024 value under different degrees of Winsorization. Panel B splits areas by mean Google Trends index level over 2023-24. Because each series is rescaled so that the area's own busiest week is 100, this is a measure of how much of the 0-100 scale the analysis years occupy, and so of how finely school-year and summer activity can be told apart, rather than a measure of search volume: its rank correlation with metropolitan population is only 0.56. Panel C reports the reliability of the two-year average used as the treatment variable, obtained from the year-to-year correlation of the Winsorized measure via the Spearman-Brown formula. Panel C is the attenuation factor applied to the bounds table in the results.\\
\end{minipage}

}

\end{table}%

\subsection{The 2025 period}\label{sec-2025}

\begin{figure}

\centering{

\pandocbounded{\includegraphics[keepaspectratio]{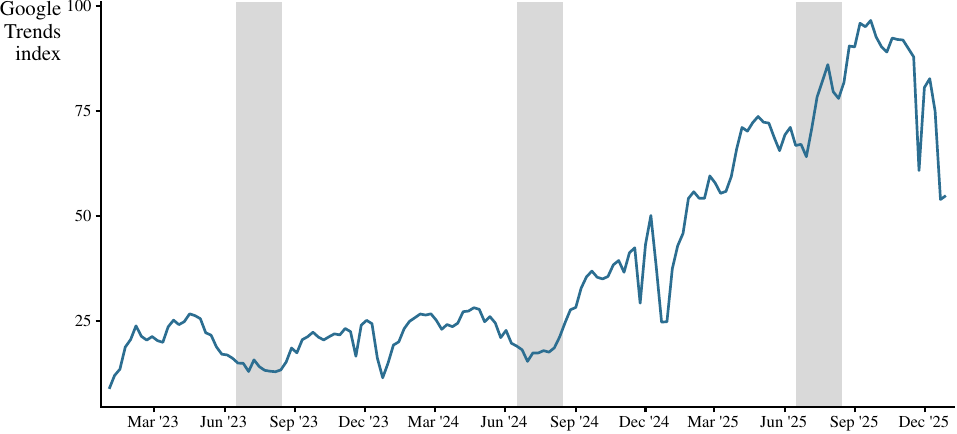}}

}

\caption{\label{fig-2025raw}Aggregate ChatGPT search interest}

\end{figure}

In the aggregate Google Trends data, the summer dip that is evident in
2023 and 2024 shows no dip at all in 2025. This might suggest the
measure stops working.\footnote{A previous version of this paper claimed
  that the measure did stop working in 2025 and used only 2023 and 2024
  data.} However, this is driven by the changing dynamics in
non-educational uses of ChatGPT in 2025, compared to 2023 and 2024.
Effectively, ChatGPT usage overall was growing so rapidly in 2025 that
the educational summer dip led to no actual aggregate dip because it was
canceled out by a strong upward trend. Figure~\ref{fig-2025raw} plots
the unweighted mean of the raw state indices. Re-estimating the
local-calendar specification of Table~\ref{tbl-calendar} on 2025 alone
gives a response of 0.135 (0.015)*** in 2025, against 0.127 (0.017)***
in 2023 and 2024. Students were still using ChatGPT on the school
calendar in 2025, it's just harder to observe against a pattern of
constantly increasing overall activity.

Each point is the unweighted mean, across the 51 state-level series, of
the weekly Google Trends index for ``chatgpt''. Shaded bands mark 16
June to 15 August, the summer window used to construct the summer dip
measure. Activity falls visibly inside the shaded band in 2023 and 2024.
In 2025 it does not: mean activity in the summer window is 75.8, against
70.2 in that year's in-session weeks, so the raw ratio is below one.

Because the dip is invisible in the aggregate series in 2025, even
though the local-calendar response is not weaker, the paper still treats
2025 as a separate case and presents results both with and without 2025.

\subsection{Remaining measurement deficiencies}\label{sec-cannot}

This section establishes that summer-to-school-year differences in
Google Trends measurements of ChatGPT activity carry a signal of K-12
educational use. The section does not establish that the measurement is
perfect. The measure cannot distinguish between heavy use by a small
number of users or widespread use across many users, and cannot
distinguish use by different kinds of students. Additionally, while
Section~\ref{sec-college} shows that college students do not drive the
results, some other form of matching seasonality may explain the summer
dip. Section~\ref{sec-limits} returns to this topic.

\section{Test score data and outcome construction}\label{sec-outcomes}

\subsection{High school: recovering continuous scores from proficiency
bands}\label{high-school-recovering-continuous-scores-from-proficiency-bands}

The primary outcome data are at the school-test-subject-grade-year level
from (SchoolDigger.com 2026), covering 67 distinct tests in 43 states,
including the PSAT, Smarter Balanced, and state examinations such as
Texas STAAR EOC, New York Regents, and Tennessee TNReady.\footnote{Raw
  SchoolDigger data is available under a non-disclosure agreement and
  cannot be redistributed. All reported results are aggregates.}

The SchoolDigger data for each test implementation is reported at the
school-grade level rather than for each student. In most cases, test
score averages are not reported, but rather the data shows the share of
students in each of up to five ordered proficiency bands.

I recover a continuous mean from the band shares in order to improve
comparability between tests.\footnote{SchoolDigger does not report band
  descriptions for most tests. For 7 of the 71 tests included in the
  data, bands are numbered from the top down. These tests, determined by
  anchoring on the reported proficiency share (which has a direction)
  are reversed. There are an additional 4 tests where orientation could
  not be determined. These tests are dropped.} I assume that scores
within a school are normally distributed with a common within-school
standard deviation. Then, the cumulative share of students who are at or
below band \(k\) along with the properties of the normal distribution
identify \(c_k - \mu_s\), where \(c_k\) is the band cut point and
\(\mu_s\) the school mean:

\begin{equation}\phantomsection\label{eq-latent}{
\Phi^{-1}\!\left(C_{sk}\right) \;=\; c_{k} - \mu_{s}
\quad\Longrightarrow\quad
\hat\mu_{s} \;=\; \frac{1}{K-1}\sum_{k}\left(\hat c_{k} - \Phi^{-1}(C_{sk})\right)
}\end{equation}

with \(\hat c_k\) set to the mean of \(\Phi^{-1}(C_{sk})\) within a test
by grade group, which fixes the location. Cut points are estimated while
pooling years, which allows year-to-year movement in scores and is valid
under the assumption that cut-points are consistent within test/year
across schools. This effectively fixes within-school test score standard
deviations at 1. This is a special case of the method in (Reardon et al.
2017) used to recover moments of continuous test score distributions
from coarsened data.

\subsection{Grades 3-8 and state-level college entrance
exams}\label{grades-3-8-and-state-level-college-entrance-exams}

Cohort-standardized district scores for grades 3 through 8 in
mathematics and reading come from the Stanford Education Data Archive
(Fahle et al. 2024), covering 2017 through 2019 and 2022 through 2024
for 9,877 districts. SEDA data on these test scores is highly detailed,
but covers a population that is unlikely to be making heavy use of AI in
their schoolwork. As such, these test scores serve as an age placebo.

State-level SAT results come from the College Board's state reports
(College Board 2026) for 2017 through 2025. State-level ACT data comes
from (ACT, Inc. 2026), beginning in 2021, when participation weight data
begins. The blended score is then the participation-weighted average of
ACT and SAT scores, using the share of students who take the SAT vs.~the
ACT, held constant over time. This data is highly limited in that it
reports at the state level, so any analysis necessarily has low power.
The analysis also shows that the data is not capable of supporting a
believable result; see Section~\ref{sec-satfail} for more data
construction details and results using this data.

\section{Empirical strategy}\label{sec-design}

\subsection{Specification}\label{specification}

The primary specification is:

\begin{equation}\phantomsection\label{eq-main}{
Y_{cy} \;=\; \sum_{y \neq 2022} \beta_{y}\,\bigl(E_{m(c)} \times \mathbf{1}[\text{year}=y]\bigr) \;+\; \alpha_{c} \;+\; \delta_{s(c),y} \;+\; \varepsilon_{cy}
}\end{equation}

where \(c\) indexes a fixed unit, school-test-subject-grade-metro, of
which there are 52,982 in the full panel and 22,117 in the balanced
panel, and a district-grade-subject-metro cell in SEDA, observed in year
\(y\).\footnote{The specification is similar in form to a
  difference-in-differences estimator with a continuous treatment. It is
  not a difference-in-differences design in the usual sense, because
  ChatGPT became available everywhere at once and no area is untreated.
  The variation being used is variation in the intensity of educational
  use, not in access.} \(m(c)\) and \(s(c)\) are, respectively, the
metropolitan area and state that unit \(c\) is situated in. \(E_{m(c)}\)
is the continuously-measured standardized ChatGPT academic usage in that
metropolitan area, averaged across 2023 and 2024 such that each
metropolitan area has a single time-invariant exposure measure.
Section~\ref{sec-timevarying} relaxes this and lets the dip vary by
metropolitan area and year, as well as incorporating the 2025 usage
measurements (the baseline pooled estimate excludes 2025, for reasons
explored in Section~\ref{sec-2025}). \(\alpha_c\) is a set of unit fixed
effects. State-by-year fixed effects \(\delta_{sy}\) absorb
state-specific recovery paths, so identification comes from comparing
metropolitan areas within a state as they move over time. 2022, the last
fully pre-ChatGPT testing year, is the omitted category.

Because testing regimes are typically set at the state level, standard
errors are clustered on state throughout the paper. One could argue that
standard errors should instead be clustered at the metro level because
that is the geographic level at which the treatment varies. None of the
qualitative results in this paper change if standard errors are instead
clustered at the metro level, although standard errors do in general
shrink. In some supplementary tests, given the relatively small number
of clusters, I use randomization inference instead of cluster-robust
standard errors, randomly reallocating metro exposure values across
metropolitan areas within the same state.

A pooled version replaces the year interactions with a single post
indicator:

\begin{equation}\phantomsection\label{eq-pooled}{
Y_{cy} \;=\; \beta\,\bigl(E_{m(c)} \times \mathrm{Post}_{y}\bigr) \;+\; \alpha_{c} \;+\; \delta_{s(c),y} \;+\; \varepsilon_{cy}
}\end{equation}

Because \(E\) is standardized across areas and outcomes are in
test-score standard deviations, the coefficient on \(E\) in each model
can be understood as representing the standard deviation change in test
scores associated with a one standard deviation increase in educational
ChatGPT usage.

\subsection{Identification}\label{identification}

Attributing a causal interpretation to the coefficient on ChatGPT usage
requires that, within a given state, metropolitan areas with high and
low levels of measured educational ChatGPT usage would have followed
similar test score recovery paths from COVID. This will not be the case
if characteristics that predict AI adoption also predict the speed of
post-COVID academic recovery.

I test for this possibility in three ways. First, I include post-COVID
pre-ChatGPT placebo years: the high school panel contains 2021 and 2022.
A nonzero coefficient in pre-ChatGPT years for the high school panel
implies deviation in recovery paths which violates the identifying
assumptions. Second, the SEDA data acts as an age placebo: grades 3
through 8 should show no effect. Third, an additional model includes
observable demographics interacted with year, so that areas differing in
composition are allowed their own recovery paths.

\subsection{Sensitivity to panel balancing}\label{sec-balancing}

\begin{figure}

\centering{

\pandocbounded{\includegraphics[keepaspectratio]{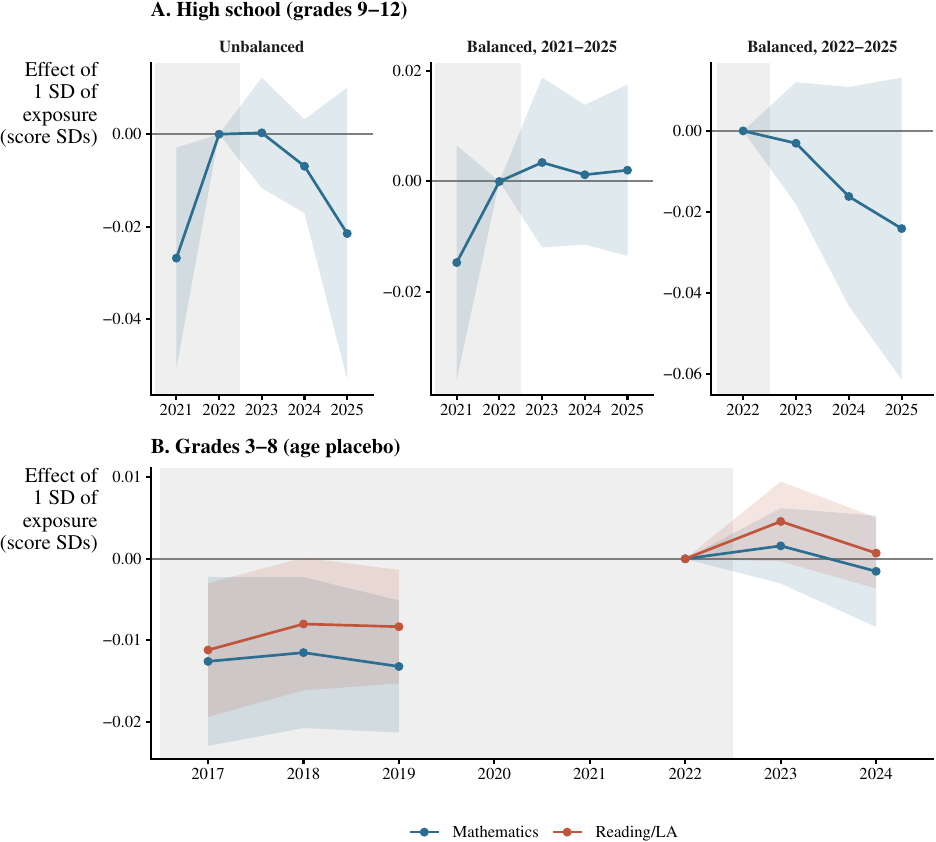}}

}

\caption{\label{fig-mainevent}Effect of ChatGPT exposure on test scores
by year}

\end{figure}

Panel A of Figure~\ref{fig-mainevent} shows the estimated event study
coefficients from Equation~\ref{eq-main} for the high school panel.
Bands are 95\% confidence intervals with standard errors, clustered on
state. All three panels use school-test-subject-grade-year level
observations and state by year fixed effects and differ only in the
sample: the full unbalanced panel, the panel balanced over 2021 through
2025, and the panel balanced over 2022 through 2025. Shading marks the
pre-ChatGPT testing years in each panel.

The first panel uses the entire high school data set. The 2021 placebo
coefficient is -0.0268 (0.0122)**. However, this nonzero result combines
together a potential assumption violation relevant to post-2022 data
with the compositional changes from 2021 to 2022 inherent in the testing
disruption; many tests occurred in 2022-2025 that were skipped in 2021.

The middle panel of Panel A in Figure~\ref{fig-mainevent} addresses
compositional changes by restricting the high school sample to the
11,811 units observed in every year from 2021 through 2025. Relative to
the unbalanced panel, this drops 11,556 units in 2021, and between
21,757 and 24,046 units in each year from 2022 through 2025.

In the 2021-2025 balanced panel, there is still a visible movement
upwards from 2021 to 2022, indicating that metropolitan areas that
would, after the launch of ChatGPT, take it up more heavily for
educational use were recovering more quickly from COVID. However, the
2021 effect is now smaller and insignificant, falling from -0.0268
(0.0122) to -0.0147 (0.0108). The apparent (but insignificant)
post-period decline that was present in the first panel is now no longer
visible, with all post-treatment years showing estimated effects very
near zero. In 2021-2025 the balanced panel, all treatment estimates are
precise zeroes.

For the rest of the paper, results use a balanced panel. However, while
the panel balanced on data from 2021-2025 is useful in that it shows a
prior trends comparison between 2021 and 2022, it heavily restricts the
data and favors states and metropolitan areas with more stable testing
regimes during COVID. As such, the remainder of the paper drops 2021
from analysis and uses a balanced panel for the years
2022-2025.\footnote{Results using the 2021-2025 balanced panel are
  available from the author and will briefly be mentioned in the results
  section.} This sample does not allow for a prior trends test, but
avoids issues related to the 2021 year, reducing compositional issues
and retaining about twice as many units (22,117).

The balanced panel is somewhat less representative, and units dropped
from the unbalanced panel were more likely to be in metropolitan areas
and states with less-stable testing regimes. This induces a geographic
distortion. The 2022-2025 balanced panel covers 149 metropolitan areas
against the unbalanced-panel 187. Retention rates vary by state, ranging
from 93\% in Massachusetts down to 10\% in Florida. For this reason,
results using the unbalanced panel will continue to be shown in
Section~\ref{sec-unbalanced}, with 2022 as reference year to reduce
reliance on the unusual 2021 comparison. The results section will
mention when unbalanced-panel results differ from balanced-panel
results; they usually do not. The unbalanced-panel results formally
contain a failed prior trends test, but this failure is driven at least
in part by compositional changes rather than a failed assumption.

Importantly, the loss of representativeness in the balanced panel
appears to be unrelated on average to treatment. The mean standardized
summer-dip measure is -0.492 before the restriction (SD 0.551) and
-0.487 (0.510) after.

\section{Results}\label{sec-results}

\subsection{Main estimates}\label{main-estimates}

The primary results were previously shown graphically in Panel A of
Figure~\ref{fig-mainevent}, with a focus on the third of its plots,
which uses the preferred panel balancing approach. This panel shows no
statistically significant effect for any post-treatment year, although
there is a general downward trend in the estimated effect. The downward
trend is one of the few results that is sensitive to the balancing
decision; if the data is further limited to testing regimes that were
also present in 2021 (middle panel) then there is no downward trend. In
either case, no year has a statistically significant effect, and even in
2025 the insignificant estimated effect is only -.02 standard
deviations.

Panel B of Figure~\ref{fig-mainevent} gives the same specification for
grades 3 through 8. Here, the pre-treatment coefficients in 2017, 2018,
and 2019 are usually statistically distinguishable from 0 at about
-0.012. However, the jump here is from pre-COVID to post-COVID (note the
gap in the graph between 2019 and 2022, where there was no data). We
would expect a change from 2019 to 2022 because of COVID. Since 2022
there is no significant or visible change, suggesting that any movement
at the high school level is not driven by geographic changes that we
would also observe in grades 3-8.

Table~\ref{tbl-main} gives estimates for estimated effects pooled across
the post-treatment years of 2023 and 2024. 2025 is omitted here because
the summer dip is not immediately apparent in the data, as in
Section~\ref{sec-2025}.

The estimated effect is shown across a number of different
specifications, including weighting by the number of test-takers, using
the share proficient instead of the estimated test score means, allowing
interactions between metro-level demographics and year, clustering on
area rather than state,\footnote{While not shown, the insignificance of
  the main pooled result is robust to using within-state randomization
  inference instead of using cluster-robust standard errors.} or adding
2025. The baseline high school estimate is -0.0096 (0.0098). Restricting
to the higher-precision areas, meaning those whose 2023-24 activity
occupies more of their own index scale, gives -0.0098 (0.0115). The
full-panel counterparts, in Table~\ref{tbl-main-unbal}, are 0.0051
(0.0051) and 0.0038 (0.0079). In both cases the result is small and
statistically insignificant.

Across the high school specifications in Panel A, one reaches
significance: allowing metropolitan areas differing in college share and
demographics their own recovery paths gives -0.0227 (0.0097)**. These
demographics include the shares of the population that are hispanic,
Black, white, who speak English very well, who have at least a HS degree
but not 4 years of college, or who have 4+ years of college, all from
the 2022 American Community Survey, as well as the college-student share
of the population, using IPEDS 2024. These are interacted with year
fixed effects. This result is significant but is fragile. The result is
sensitive to the balancing decision, and does not survive using a
time-varying treatment (as in Section~\ref{sec-timevarying}), or
restricting the sample to the metropolitan areas where the measure is
most precise.\footnote{The estimated effect is 0.0026 (0.0059) in the
  full unbalanced panel, -0.0026 (0.0053) in the model variation
  allowing for time-varying ChatGPT usage, and -0.0152 (0.0132) if
  limiting the sample to the higher-precision metropolitan areas.}

Purging college share from the treatment itself, which is the direct way
to address college contamination, leaves the estimate at -0.0143
(0.0089).

In the unbalanced version in Table~\ref{tbl-main-unbal}, where the
demographic-trend specification is not significant, local-calendar
exposure is significant, at 0.0176 (0.0090)* in the unbalanced panel
against 0.0027 (0.0129). Local-calendar exposure uses state-calendar
summer definitions instead of a shared one, after removing
national-level seasonality trends, similar to the test in
Table~\ref{tbl-calendar}. This is the one result in that panel that
might be read as positive evidence. However, it is not robust: it is one
significant result out of 12, risking specification search issues, and
the balanced panel loses significance not just because the smaller
sample increases standard errors, but also because the estimated
coefficient shrinks considerably.

\begin{table}

\caption{\label{tbl-main}Effect of ChatGPT exposure on test scores}

\centering{

\fontsize{7.0pt}{8.0pt}\selectfont
\begin{tabular*}{\linewidth}{@{\extracolsep{\fill}}lrr}
\toprule
Specification & Estimate & Obs \\ 
\midrule\addlinespace[2.5pt]
\multicolumn{3}{l}{A. High school (grades 9-12), balanced 2022-2025 panel} \\[2.5pt] 
\midrule\addlinespace[2.5pt]
Baseline (all metros) & -0.0096 (0.0098) & 66,351 \\ 
Higher-precision metros only & -0.0098 (0.0115) & 51,222 \\ 
Weighted by test-takers & 0.0008 (0.0097) & 62,183 \\ 
Unwinsorised exposure & -0.0123 (0.0111) & 66,351 \\ 
Exposure purged of college share & -0.0143 (0.0089) & 66,351 \\ 
Local-calendar exposure & 0.0027 (0.0129) & 66,351 \\ 
Proficiency-share outcome & -0.0015 (0.0039) & 66,351 \\ 
Metro covariates x year & -0.0227 (0.0097)** & 65,802 \\ 
State + year FE (no state x year) & 0.0090 (0.0253) & 66,351 \\ 
Cluster on metro & -0.0096 (0.0089) & 66,351 \\ 
Include 2025 & -0.0144 (0.0126) & 88,468 \\ 
\midrule\addlinespace[2.5pt]
\multicolumn{3}{l}{B. Grades 3-8 (age placebo)} \\[2.5pt] 
\midrule\addlinespace[2.5pt]
Mathematics & -0.0006 (0.0022) & 116,253 \\ 
Reading/LA & 0.0030 (0.0019) & 124,319 \\ 
Mathematics, + SES x year & -0.0001 (0.0022) & 111,495 \\ 
Reading/LA, + SES x year & 0.0041 (0.0020)* & 119,052 \\ 
\midrule\addlinespace[2.5pt]
\multicolumn{3}{l}{C. High school, time-varying exposure measure} \\[2.5pt] 
\midrule\addlinespace[2.5pt]
Baseline (all metros) & 0.0022 (0.0036) & 88,441 \\ 
Higher-precision metros only & 0.0058 (0.0033)* & 68,296 \\ 
Weighted by test-takers & 0.0024 (0.0039) & 82,813 \\ 
Unwinsorised exposure & -0.0003 (0.0040) & 88,441 \\ 
Proficiency-share outcome & 0.0017 (0.0015) & 88,441 \\ 
Metro covariates x year & -0.0026 (0.0053) & 87,733 \\ 
State + year FE (no state x year) & 0.0086 (0.0140) & 88,441 \\ 
Cluster on metro & 0.0022 (0.0035) & 88,441 \\ 
Drop 2025 (weak exposure measure) & -0.0061 (0.0080) & 66,324 \\ 
\midrule\addlinespace[2.5pt]
\multicolumn{3}{l}{D. High school, time-varying exposure by year} \\[2.5pt] 
\midrule\addlinespace[2.5pt]
2022 (reference) & 0 &  \\ 
2023 & 0.0100 (0.0066) &  \\ 
2024 & -0.0017 (0.0046) &  \\ 
2025 & 0.0172 (0.0160) &  \\ 
\bottomrule
\end{tabular*}
\begin{minipage}{\linewidth}
Notes: Coefficients are test-score standard deviations per standard deviation of exposure, except the proficiency-share rows, which are shares on a 0-1 scale. All specifications carry unit and state-by-year fixed effects. Panel A is the balanced 2022-2025 high school panel over 2022-2024; Panel B is grades 3-8, unrestricted; Panels C and D replace the fixed exposure measure with the area-by-year dip over 2022-2025, normalized to zero in 2022. Metro covariates are the seven area characteristics listed in the text. Unbalanced-panel versions of every panel are in the appendix. Significance: * p<0.10, ** p<0.05, *** p<0.01.\\
\end{minipage}

}

\end{table}%

The grades 3 through 8 placebo does not show signs that threaten the
main high school design. The estimates are -0.0006 (0.0022) for
mathematics and 0.0030 (0.0019) for reading. Allowing district COVID
recovery paths to vary by socioeconomic status does lead to a
statistically significant effect for reading, but only at the 10\%
level, and the estimated effect is only .004 standard deviations.

\subsection{Effect heterogeneity}\label{effect-heterogeneity}

While the overall effect is null, it is possible that educational
ChatGPT usage has test score impacts for some subset of the tests or
schools examined.

Panel A of Table~\ref{tbl-hetero} reports estimated effects separately
by grade and subject for all combinations for which at least five states
can be included in the sample. One of the 11 cells is significant at the
5\% level, science end-of-course at -0.0260 (0.0097)** on 7 states, and
mathematics end-of-course reaches the 10\% level. This is some evidence
of an effect specifically in Science end-of-course tests, although this
is a case where there is one significant effect among many different
subsamples, and the effects vary widely across subsamples, indicating
that this result may just be noise.

Beyond grade level and subject, an important potential element of effect
heterogeneity is how the effect might differ across the performance
distribution. (Yu et al. 2024) and (Hausman, Rigbi, and Weisburd 2025)
both find compression effects of AI usage, with gains concentrated at
the bottom of the distribution. Results across the performance
distribution are shown in Panels B and C of Table~\ref{tbl-hetero}.

I first look at the effect across school/test combinations at different
levels of the performance distribution. Splitting school/test
combinations by their own pre-period performance within the balanced
panel gives 0.0237 (0.0108)** for the bottom third of schools, 0.0233
(0.0126)* for the middle and -0.0313 (0.0100)*** for the top. Schools
that initially performed poorly see positive impacts of ChatGPT
exposure, those in the middle see less-positive impacts, and those at
the top see negative impacts, keeping in mind that this tracks variation
in educational ChatGPT usage within those bins and so cannot be
explained by differences between the terciles in usage. The difference
between the top and bottom terciles is itself significant. Estimating a
fully-interacted model and testing for the difference between
coefficients, the bottom-minus-top difference is 0.0550 (0.0191), which
rejects equality at the 1\% level with standard errors clustered on
state (\(p = 0.007\), on 31 clusters). If instead we use randomization
inference to test the difference, it is significant at the 5\% level
(\(p = 0.010\)).

This is the most notable non-null result in this paper, but it is not
robust. The terciles are formed by pre-treatment measurement of the
outcome variable, so regression to the mean may be an issue. This result
is also sensitive to panel balancing. The unbalanced panel results in
Section~\ref{sec-unbalanced} show the most positive effects for the
middle tercile, instead of compression, and the bottom-minus-top
difference is insignificant. This is also a case where the use of the
2021-2025 balanced panel leads to a different result: the 2021-2025
balanced panel shows a similar pattern to the unbalanced panel, with the
middle tercile showing the largest effect, without a significant
difference between bottom and top. This result is interesting, and is
consistent with the literature, but is not robust in these data.

The result comparing school terciles, however, looks at ChatGPT effects
at different parts of the school performance distribution. The previous
literature more directly applies to a prediction of how \emph{students}
at the bottom of the performance distribution might perform, compared to
students at the top. To evaluate this, instead of looking at the impact
of educational ChatGPT usage on average scores, I limit the data to
tests that, in the original SchoolDigger data, report at least four
different performance bands. Then I look at the share of students who
score above the lowest performance bin for each test, and the share of
students who score in the highest. A one-standard-deviation increase in
the share of educational ChatGPT usage changes the share of students
scoring above the lowest band by -0.0045 (0.0048) and the share scoring
in the highest band by -0.0024 (0.0014). Neither reaches the 5\% level.
The top-band estimate reaches the 10\% level, although with an estimate
that is smaller in absolute terms than for the lowest-band margin.

Taking both the school and student results together, there is some
evidence that ChatGPT effects are worse for higher-performing students
and schools, but this is a weak result.

\begin{table}

\caption{\label{tbl-hetero}Effect heterogeneity}

\centering{

\fontsize{7.0pt}{8.0pt}\selectfont
\begin{tabular*}{\linewidth}{@{\extracolsep{\fill}}lrrrr}
\toprule
 & Estimate & p-value & States & Obs \\ 
\midrule\addlinespace[2.5pt]
\multicolumn{5}{l}{A. By subject and grade} \\[2.5pt] 
\midrule\addlinespace[2.5pt]
English, 9th Grade & -0.0319 (0.0396) & 0.465 & 5 & 1,371 \\ 
English, 10th Grade & 0.0177 (0.0119) & 0.163 & 12 & 4,125 \\ 
English, 11th Grade & -0.0070 (0.0147) & 0.644 & 13 & 8,028 \\ 
English, End of course & 0.0202 (0.0163) & 0.271 & 6 & 9,042 \\ 
Mathematics, 9th Grade & 0.0324 (0.0590) & 0.613 & 5 & 2,100 \\ 
Mathematics, 10th Grade & 0.0150 (0.0103) & 0.175 & 11 & 2,823 \\ 
Mathematics, 11th Grade & -0.0145 (0.0141) & 0.325 & 14 & 8,067 \\ 
Mathematics, End of course & -0.0258 (0.0135)* & 0.094 & 9 & 11,262 \\ 
Science, 10th Grade & -0.0140 (0.0080) & 0.138 & 6 & 1,419 \\ 
Science, 11th Grade & 0.0042 (0.0149) & 0.782 & 11 & 5,397 \\ 
Science, End of course & -0.0260 (0.0097)** & 0.037 & 7 & 10,857 \\ 
\midrule\addlinespace[2.5pt]
\multicolumn{5}{l}{B. Schools, by baseline performance tercile} \\[2.5pt] 
\midrule\addlinespace[2.5pt]
Bottom third & 0.0237 (0.0108)** & 0.036 &  & 22,119 \\ 
Middle third & 0.0233 (0.0126)* & 0.074 &  & 22,116 \\ 
Top third & -0.0313 (0.0100)*** & 0.004 &  & 22,116 \\ 
Bottom minus top & 0.0550 (0.0191) & 0.007 &  &  \\ 
\midrule\addlinespace[2.5pt]
\multicolumn{5}{l}{C. Students, within-school distribution} \\[2.5pt] 
\midrule\addlinespace[2.5pt]
Share above lowest band & -0.0045 (0.0048) & 0.351 &  & 65,049 \\ 
Share in highest band & -0.0024 (0.0014)* & 0.096 &  & 65,048 \\ 
\bottomrule
\end{tabular*}
\begin{minipage}{\linewidth}
Notes: Estimates on the balanced 2022-2025 panel over 2022-2024, with unit and state by year fixed effects and standard errors clustered on state. Each row of Panel A is a separate estimate of Equation 4 restricted to that subject and grade; States is the number of clustering states behind that row, and cluster-robust errors are less trustworthy where it is small. Panel B splits school-test combinations into terciles of their own pre-period performance and estimates each separately; coefficients in Panels A and B are in test-score standard deviations per standard deviation of exposure. A comparison of the top and bottom tercile coefficients using randomization inference gives p = 0.010 . Panel C is restricted to tests reporting at least four proficiency bands, and its outcomes are shares of students on a 0-1 scale rather than test-score standard deviations. The full-panel versions of all three panels are in the appendix. Significance: * p<0.10, ** p<0.05, *** p<0.01.\\
\end{minipage}

}

\end{table}%

\subsection{Time-varying educational ChatGPT
usage}\label{sec-timevarying}

Every estimate reported so far gives a metropolitan area a single
exposure value that applies to the whole panel, as specified in
Section~\ref{sec-design}. This reduces noise in the summer-dip measure,
assuming that the dip is a durable regional characteristic, for which
there is some support in Section~\ref{sec-reliability}. However, this
will miss if a given region's test scores respond to time variation in
their educational ChatGPT usage, and also in most specifications omits
the 2025 summer dip measurement.

This section measures the dip separately by metropolitan area and by
year. The summer dip in 2022 is set to zero, and each year's dip is
centered on its own mean, so there is no common level shift across
years. All years are then divided by a common 2023-2025 standard
deviation. The resulting measure has a standard deviation of 0.65 in
2023, 0.72 in 2024 and 0.21 in 2025. The reduced standard deviation in
2025 is explained by the descriptives in Section~\ref{sec-2025}.

The specification using the metro- and year-varying summer dip measure
\(E_{m(c)y}\) re-uses the original specification to allow the
coefficient to vary by year, and the pooled version replaces
\(E_{m(c)} \times \mathbf{1}[\text{year}=y]\) with \(E_{m(c)y}\).

Panel D of Table~\ref{tbl-main} gives the event study on the balanced
panel. The coefficients are 0.0100 (0.0066) in 2023, -0.0017 (0.0046) in
2024 and 0.0172 (0.0160) in 2025. None of these estimated effects are
statistically distinguishable from zero. The 2025 estimate is
meaningfully large and positive, but is noisily measured and
insignificant.

Pooled estimates using different specifications are in Panel C of
Table~\ref{tbl-main}. No specification in the table is significant at
the 5\% level. The full panel results, shown in Panels C and D of
Table~\ref{tbl-main-unbal}, are meaningfully different. In the
unbalanced panel, the time-varying specification gives 0.0078 (0.0036)**
across all metropolitan areas and 0.0145 (0.0035)*** restricting to
higher-precision areas. However, this result is not robust. It both
depends on composition changes (there is no effect in the 2021-2025
balanced panel), and is driven specifically by 2025, where the measure
is weakest. The unbalanced-panel pooled result using 2023-2024 only is
-0.0007 (0.0047).

\subsection{State-level college entrance
exams}\label{state-level-college-entrance-exams}

It is possible with this design to estimate the impact of educational
ChatGPT usage on college entrance exams. This data is only available at
the state level. Because the analysis is state-level, the statistical
power of this analysis is low. Additionally, a pre-trends analysis
fails. These results also show no impact of educational ChatGPT usage,
but are not trustworthy. For completeness, these results are shown in
Section~\ref{sec-satfail}.

\subsection{Measurement error}\label{sec-bounds}

The results so far have shown insignificant results, and the cases where
the results are statistically significant have not been robust. However,
it is also the case that the primary measure is based on Google Trends
data, which is only a representation of the actual measurement we want,
actual ChatGPT usage for high school education purposes. Despite
attempts to validate the measure throughout Section~\ref{sec-measure},
estimated zero effects may reflect the ChatGPT usage signal being lost
by being noisily measured by Google Trends activity. I argue that
Section~\ref{sec-measure} establishes that the Google Trends-measured
summer dip carries a useful signal, but there is also noise. In this
section I produce estimate bounds that allow for measurement error,
under the simplifying assumption that the error is classical in nature.

Table~\ref{tbl-bounds} widens each interval by the reliability of the
exposure measure from Section~\ref{sec-reliability}, the classical
attenuation factor for a mismeasured regressor. This treats the 2023 and
2024 dips as two measures of the same true value with noise assumed to
be additive and independent. Neither assumption holds exactly. The
measure is a bounded, Winsorized ratio, so additivity fails, indicating
that the adjustment is too small. However, true usage changed between
2023 and 2024, so independence fails in a way that understates
reliability (since true-value changes appear as reduced reliability),
indicating the adjustment is too large. The exercise is therefore a
check on whether a basic check on an attenuation correction appears to
be large enough to change the conclusion, not a claim that the results
are properly adjusted.

\begin{table}

\caption{\label{tbl-bounds}Bounds on the effect of educational ChatGPT
exposure}

\centering{

\fontsize{7.0pt}{8.0pt}\selectfont
\begin{tabular*}{\linewidth}{@{\extracolsep{\fill}}lrlrr}
\toprule
Sample & Estimate & 95\% CI & Bound & Bound, corrected \\ 
\midrule\addlinespace[2.5pt]
High school, all subjects & -0.0096 & [-0.0297, 0.0105] & 0.0297 & 0.0372 \\ 
High school, higher-precision metros & -0.0098 & [-0.0334, 0.0138] & 0.0334 & 0.0361 \\ 
Grades 3-8, mathematics & -0.0006 & [-0.0051, 0.0039] & 0.0051 & 0.0064 \\ 
Grades 3-8, reading & 0.0030 & [-0.0009, 0.0069] & 0.0069 & 0.0086 \\ 
\bottomrule
\end{tabular*}
\begin{minipage}{\linewidth}
Notes: Estimates and intervals are in test-score standard deviations per standard deviation of exposure. The bound column is the larger absolute end of the 95\% confidence interval. The corrected column divides by the classical attenuation factor for classical measurement error in a regressor. The high school rows use the balanced panel; the full-panel bounds are in the appendix.\\
\end{minipage}

}

\end{table}%

A reasonable conclusion that the data in this paper can support, taking
these findings only as an indication that the proper measurement
correction would not be enormous, is that a one standard deviation
increase in educational ChatGPT usage across metropolitan areas changed
high school test scores by less than about 0.04 standard deviations in
absolute terms, based on Table~\ref{tbl-bounds}.

\section{What this design can and cannot detect}\label{sec-limits}

The extent of the available data, as well as the coincident timing of
ChatGPT with COVID, limit what can be learned in this context. The
results in this paper are supported to the extent they can be, but there
are limits in the claims that can be made. Several limitations are
structural, affecting how strongly the results should be taken.

First, the treatment is a share. The summer dip measures what fraction
of an area's ChatGPT activity is school-timed, not how much any student
uses it. An area in which 50\% of students and 50\% of non-students
regularly use ChatGPT would report a similar summer dip to an area in
which 30\% of students and 30\% of non-students regularly use ChatGPT
(if the two areas also have similar population shares of students).
There is variation across regions in absolute usage levels that would
make this relevant. In 2024, average monthly visits per capita (not
unique visitors per capita) for the fifteen states for which we have
Similarweb traffic data ranged from 0.82 in Ohio to 1.44 in California.
Ohio is a heavy outlier here, and the 25th and 75th percentiles of
per-capita visits are 1.00 and 1.25. So the variation in usage is not
extreme but is relevant and may impact results.

Second, while Section~\ref{sec-calendar} establishes that the
time-series movement in aggregate activity is driven by local K-12
calendars, there is still the possibility of residual seasonal
confounding from other ChatGPT usage sources that match K-12 timing. The
most obvious of these is college students. Section~\ref{sec-college}
quantifies their contribution and shows it is bounded, and
Table~\ref{tbl-main} shows that the result is robust to residualizing
the treatment variable with respect to the college student share.
However, other people, if their ChatGPT usage patterns change in the
summer, can threaten identification. Resolving this issue would require
usage data with age and content attached, for instance provider-side
data of the kind analyzed by (Chatterji et al. 2025) but disaggregated
geographically.

Third, no individual student is observed. If AI substantially helps some
students and harms others, aggregate scores can be flat. Similarly, if
the impact is concentrated in a small share of students, the overall
effect may be close to zero even if that subset of students is greatly
affected. The distributional results in Section~\ref{sec-results}
provide some evidence against large offsetting effects, but they only
have limited ability to estimate effects separately by student,
especially if the features that drive the effect are unrelated to
initial test performance.

Fourth, there is no way to fully disentangle these short-term effects
from those of COVID. The greatest impacts of COVID on education were in
2021 and the study is able to separate out that year, but the relevant
window remains one in which schools were on COVID recovery paths.
Because of when ChatGPT was introduced, and since it was introduced
universally at a single time point, this issue will affect any
observational estimate of ChatGPT impacts.

Finally, while much of the paper is dedicated to validating the
measurement of educational ChatGPT usage, it remains a proxy for actual
usage rates among students. This would be less of an issue if results
were large and significant, since we might expect that in this case a
proxy measurement would attenuate results toward zero.

\section{Conclusion}\label{conclusion}

Educational ChatGPT use, as measured by the concentration of an area's
ChatGPT activity within the school year, has no detectable effect on
high school test scores in 2023 and 2024. The estimate is
indistinguishable from zero across subjects, grades, and nearly all
specification and data-limitation variants. Statistically significant
results tend not to be robust, and estimates are precise enough that
effects larger than about .04 standard deviations of test scores per
standard deviation of exposure are excluded.

These results appear contrary to two separate strains of thought and
results about AI and learning. First, unlike many of the studies finding
that intentional in-class usage of ChatGPT or specialized AI tools show
significant learning gains in experimental environments, this study
shows no positive impact. Second, unlike studies like (Strömberg, Lei,
and Wu 2026), which find negative learning effects of individual AI
adoption or in courses exposed to AI, or in the general concerns from
educators about the impact of unsupervised AI usage, this study shows no
negative impact. Relevant here are the differences in what the treatment
is (unsupervised usage vs.~a specialized tool) and the scale on which
the effect is measured (here, aggregate, as opposed to per-student).

It is perhaps not surprising that unsupervised usage fails to find a
positive impact where more intentional usage or specialized AI tools
show positive impacts. The lack of a negative aggregate effect is more
surprising, and suggests areas for future work. If, as in (Strömberg,
Lei, and Wu 2026) or (Chirikov 2026), AI may improve grades without
improving learning, we would expect negative impacts in this study. Some
potential implications of an aggregate null effect in this study might
include: (a) the true effect of unsupervised ChatGPT usage on learning
may be zero, not negative, (b) the effect is different in high school
than in college, (c) whatever per-student impact there may be, those
impacts are small enough when weighted by usage that they do not impact
score aggregates, (d) the kinds of learning-avoidant AI usage that those
studies isolate are more rare than expected among AI users, or (e) there
are corrective forces related to these tests, where if there is a
negative ChatGPT impact, it is small or malleable enough that educators
interested in improving scores can overcome that negative impact,
leading to no effect overall.

A sixth explanation, (f), might be that the individual effects, weighted
by student usage, were not large enough to impact aggregates by 2024,
but the impact may grow over time. This would be consistent with
(Strömberg, Lei, and Wu 2026), and there was some suggestive but
weakly-measured evidence in this paper that the effect was larger in
2025 than in 2023-2024 that would support this interpretation.
Evaluating this possibility as more post-ChatGPT testing data becomes
available is then important. Finally, (g) the extent of ChatGPT usage
among high school students may be so strong that the introduction of
ChatGPT affects performance nationally, but the geographic variation in
usage levels, which is the identifying variation in this paper, is
simply not large enough to matter.

It would be premature to conclude, based on this study, that
unsupervised AI usage has no learning impacts. Rather, the message is
that whatever these impacts are, we can estimate that they are small
enough to not meaningfully move aggregate outcomes, or at least have
been that small thus far for high school students. The exact reason for
this remains open. These findings do not mean that unsupervised AI usage
poses no threat to learning, but it is a result that can provide a
measured relief for educators deeply concerned about the impact of this
new technology on learning.

\section{Declaration of Generative AI and AI-assisted Technologies in
the Manuscript Preparation
Process}\label{declaration-of-generative-ai-and-ai-assisted-technologies-in-the-manuscript-preparation-process}

During the preparation of this work, the author used Claude Code with
Opus 4.6 and Opus 5 for assistance with coding, constructing and
validating the school-calendar measurement tests, searching for
additional papers for the literature review, and making suggestions for
revision. All ideas, first-draft writing, final review, and edits come
from the author, who is fully responsible for the content.

\section*{References}\label{references}
\addcontentsline{toc}{section}{References}

\phantomsection\label{refs}

\FloatBarrier

\newpage
\appendix
\renewcommand{\thesection}{\Alph{section}}
\setcounter{section}{0}
\counterwithin{figure}{section}
\counterwithin{table}{section}
\renewcommand{\section}[1]{\refstepcounter{section}%
\begin{Large}\bfseries Appendix \thesection: #1\end{Large}}

\section{Supporting tables and
figures}\label{supporting-tables-and-figures}

\FloatBarrier

\begin{table}

\caption{\label{tbl-calgroups}K-12 calendar timing groups}

\centering{

\fontsize{7.0pt}{8.0pt}\selectfont
\begin{tabular*}{\linewidth}{@{\extracolsep{\fill}}lrl}
\toprule
Window & States & Examples \\ 
\midrule\addlinespace[2.5pt]
\multicolumn{3}{l}{Start of school year} \\[2.5pt] 
\midrule\addlinespace[2.5pt]
Early August & 11 & Alabama, Arizona, Georgia \\ 
Mid-August & 14 & Arkansas, California, Colorado \\ 
Late August & 15 & Connecticut, Delaware, District of Columbia \\ 
Early September & 11 & Alaska, Maine, Massachusetts \\ 
\midrule\addlinespace[2.5pt]
\multicolumn{3}{l}{End of school year} \\[2.5pt] 
\midrule\addlinespace[2.5pt]
Mid-May & 4 & Arizona, Missouri, Nebraska \\ 
Late May & 22 & Alabama, Arkansas, Colorado \\ 
Early June & 11 & Alaska, California, Connecticut \\ 
Mid-June & 10 & Delaware, District of Columbia, Illinois \\ 
Late June & 4 & Maine, New York, Rhode Island \\ 
\bottomrule
\end{tabular*}
\begin{minipage}{\linewidth}
Notes: States are assigned to half-month windows for the start and end of the school year, from Clell Wade Coaches Directory.\\
\end{minipage}

}

\end{table}%

\begin{table}

\caption{\label{tbl-loo}Leave-one-timing-group-out estimates}

\centering{

\fontsize{7.0pt}{8.0pt}\selectfont
\begin{tabular*}{\linewidth}{@{\extracolsep{\fill}}lrr}
\toprule
Dropped & Estimate & States \\ 
\midrule\addlinespace[2.5pt]
\multicolumn{3}{l}{Fall window (drop a start-date group)} \\[2.5pt] 
\midrule\addlinespace[2.5pt]
Early August & 0.149 (0.022)*** & 40 \\ 
Early September & 0.092 (0.019)*** & 40 \\ 
Mid-August & 0.128 (0.019)*** & 37 \\ 
Late August & 0.133 (0.022)*** & 36 \\ 
\midrule\addlinespace[2.5pt]
\multicolumn{3}{l}{June window (drop an end-date group)} \\[2.5pt] 
\midrule\addlinespace[2.5pt]
Late May & 0.057 (0.022)*** & 29 \\ 
Early June & 0.087 (0.019)*** & 40 \\ 
Mid-May & 0.075 (0.019)*** & 47 \\ 
Mid-June & 0.081 (0.013)*** & 41 \\ 
Late June & 0.088 (0.026)*** & 47 \\ 
\bottomrule
\end{tabular*}
\begin{minipage}{\linewidth}
Notes: Each row drops all states in one calendar timing group and re-estimates Equation 1, confirming that no single group of states drives the result. Coefficients are shares of the area-year mean index, as in the calendar table. Standard errors clustered on state in parentheses. Significance: * p<0.10, ** p<0.05, *** p<0.01.\\
\end{minipage}

}

\end{table}%

\begin{figure}

\centering{

\pandocbounded{\includegraphics[keepaspectratio]{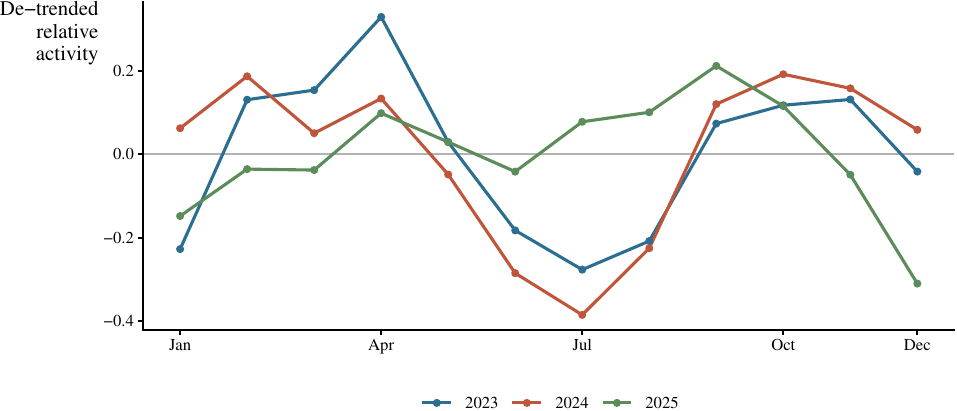}}

}

\caption{\label{fig-seasonality}ChatGPT search interest by month and
year}

\end{figure}%

Each state-year series is divided by its own mean and a within-year
linear trend is removed separately for each year, so that the seasonal
shape can be compared across years of very different activity levels.
Removing a single trend from pooled years produces artifacts that can be
mistaken for seasonality. The summer decline appears in 2023 and 2024.
The March and April peak in 2023 is not a school effect: ChatGPT's
largest capability releases fell in that window, including GPT-4 on 14
March 2023. National shocks of this kind are absorbed by the
week-by-year fixed effects in Equation~\ref{eq-calendar}, which is why
the identification strategy does not rely on the shape of the national
seasonal profile.

\begin{figure}

\centering{

\pandocbounded{\includegraphics[keepaspectratio]{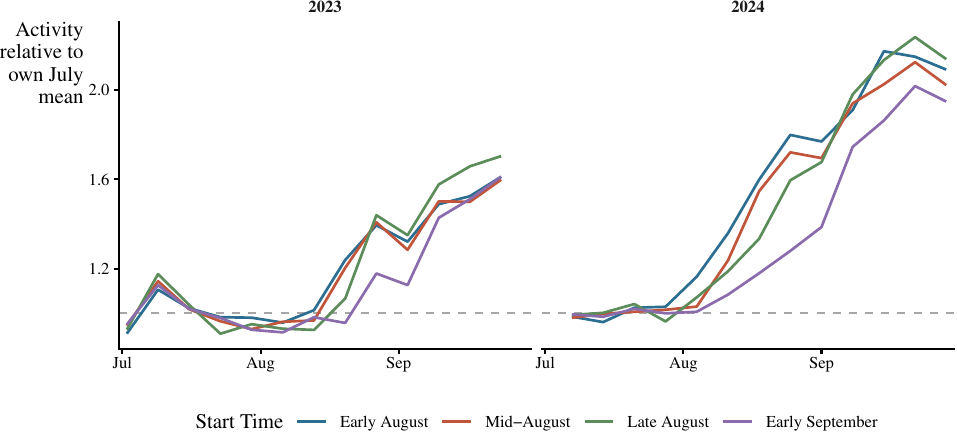}}

}

\caption{\label{fig-profile}ChatGPT activity rebounds earlier where
school starts earlier}

\end{figure}%

\begin{figure}

\centering{

\pandocbounded{\includegraphics[keepaspectratio]{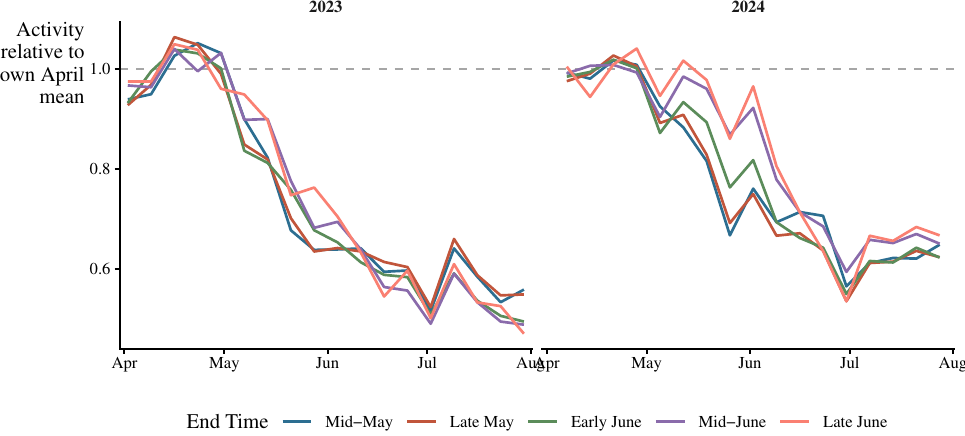}}

}

\caption{\label{fig-profile-end}ChatGPT activity ends earlier where
school ends earlier}

\end{figure}%

\begin{table}

\caption{\label{tbl-sates}State-level SAT estimates by year}

\centering{

\fontsize{7.0pt}{8.0pt}\selectfont
\begin{tabular*}{\linewidth}{@{\extracolsep{\fill}}rrrr}
\toprule
Year & SAT ERW & SAT Math & SAT Total \\ 
\midrule\addlinespace[2.5pt]
2017 & 0.114 (0.076) & 0.070 (0.074) & 0.092 (0.075) \\ 
2018 & 0.110 (0.042) & 0.102 (0.044) & 0.107 (0.042) \\ 
2019 & 0.103 (0.037) & 0.104 (0.036) & 0.103 (0.036) \\ 
2021 & 0.074 (0.035) & 0.045 (0.031) & 0.058 (0.033) \\ 
2022 & 0.000 (0.000) & 0.000 (0.000) & 0.000 (0.000) \\ 
2023 & 0.047 (0.065) & 0.028 (0.054) & 0.040 (0.060) \\ 
2024 & 0.021 (0.047) & 0.008 (0.039) & 0.014 (0.042) \\ 
2025 & 0.022 (0.056) & -0.008 (0.060) & 0.006 (0.058) \\ 
\bottomrule
\end{tabular*}
\begin{minipage}{\linewidth}
Notes: Coefficients are in test-score standard deviations per standard deviation of state exposure, relative to 2022, with state and year fixed effects. High-exposure states sit well above their 2022 level throughout 2017-2019, declining monotonically toward the omitted year, which is the pre-trend that disqualifies this design. 2020 is excluded because of pandemic testing disruption. Standard errors clustered on state in parentheses.\\
\end{minipage}

}

\end{table}%

\begin{table}

\caption{\label{tbl-sedagrade}Grades 3-8 effects by grade}

\centering{

\fontsize{7.0pt}{8.0pt}\selectfont
\begin{tabular*}{\linewidth}{@{\extracolsep{\fill}}rrr}
\toprule
Grade & Estimate & Obs \\ 
\midrule\addlinespace[2.5pt]
3 & -0.0039 (0.0044) & 42,533 \\ 
4 & 0.0067 (0.0028)** & 43,146 \\ 
5 & -0.0014 (0.0071) & 43,036 \\ 
6 & -0.0015 (0.0035) & 42,311 \\ 
7 & 0.0052 (0.0060) & 38,311 \\ 
8 & 0.0044 (0.0031) & 31,235 \\ 
\bottomrule
\end{tabular*}
\begin{minipage}{\linewidth}
Notes: Coefficients are in test-score standard deviations per standard deviation of exposure, using 2022 as the pre-period, with district by grade by subject and state by year fixed effects. If the design were picking up genuine ChatGPT use rather than a regional confounder, effects should be larger in grades 7 and 8 than in grades 3 and 4; they are not systematically ordered. Standard errors clustered on state in parentheses. Significance: * p<0.10, ** p<0.05, *** p<0.01.\\
\end{minipage}

}

\end{table}%

\FloatBarrier

\section{Data construction and combination details}\label{sec-merging}

Test score data are linked to Google Trends metropolitan areas, which
are defined similarly to designated market areas and are coarser than
the metropolitan definitions used by SEDA or by the Census. SEDA
supplies a school-to-MSA crosswalk, which I use to link SEDA data to
Google Trends. SchoolDigger schools are linked by ZIP code to Core-Based
Statistical Areas via a commercial crosswalk (Zip Code to MSA Mapping
2026) and then to Google Trends areas by string matching followed by
manual checking.

Area-level demographic controls come from the 2022 American Community
Survey (Ruggles et al. 2025) (from before the launch of ChatGPT).
College student share comes from IPEDS (U.S. Department of Education,
National Center for Education Statistics 2025). I take 2024 IPEDS
college student counts (undergraduate, graduate, and sub-baccalaureate,
subtracting out the number of students who are exclusively in distance
education) for all colleges in a given metro (CBSA). This is compared to
the census-derived population ZCTA values for all zip codes in the CBSA.

\begin{table}

\caption{\label{tbl-coverage}Analysis samples}

\centering{

\fontsize{7.0pt}{8.0pt}\selectfont
\begin{tabular*}{\linewidth}{@{\extracolsep{\fill}}lrrrr}
\toprule
Year & Observations & Schools or districts & Areas & States \\ 
\midrule\addlinespace[2.5pt]
\multicolumn{5}{l}{A. High school (grades 9-12), SchoolDigger} \\[2.5pt] 
\midrule\addlinespace[2.5pt]
2021 & 23,367 & 7,641 & 175 & 39 \\ 
2022 & 33,568 & 9,015 & 176 & 41 \\ 
2023 & 35,293 & 9,443 & 182 & 42 \\ 
2024 & 35,687 & 9,354 & 180 & 41 \\ 
2025 & 35,857 & 8,985 & 183 & 40 \\ 
\midrule\addlinespace[2.5pt]
\multicolumn{5}{l}{B. High school (grades 9-12), balanced 2022-2025} \\[2.5pt] 
\midrule\addlinespace[2.5pt]
2022-2025 (each year) & 22,117 & 6,967 & 149 & 31 \\ 
\midrule\addlinespace[2.5pt]
\multicolumn{5}{l}{C. Grades 3-8, SEDA} \\[2.5pt] 
\midrule\addlinespace[2.5pt]
2017 & 92,460 & 9,363 & 178 &  \\ 
2018 & 92,972 & 9,307 & 177 &  \\ 
2019 & 94,805 & 9,337 & 177 &  \\ 
2022 & 79,885 & 7,744 & 169 &  \\ 
2023 & 81,395 & 7,824 & 172 &  \\ 
2024 & 79,292 & 7,736 & 168 &  \\ 
\bottomrule
\end{tabular*}
\begin{minipage}{\linewidth}
Notes: The high school panels are school-test-subject-grade-year observations with a recovered continuous mean. Panel B restricts Panel A to the units observed in every year from 2022 through 2025, which is the sample used for the main results; being balanced, it carries identical counts in each of those years, so it is reported as a single row. Panel C is district by grade by subject observations from SEDA. All are linked to metropolitan-level ChatGPT exposure.\\
\end{minipage}

}

\end{table}%

Table~\ref{tbl-coverage} includes the full data set, but the high school
panel is not balanced across years. Because of COVID disruptions,
especially in 2021, many tests are not available in that year, and also
SchoolDigger has added additional tests over time. Of the 52,982 units
(163,772 observations, 10,068 schools, 43 states) in the full panel,
11,811 units (59,055 observations, 4,542 schools, 24 states) are
observed in all five years from 2021 through 2025, and 22,117 units
(88,468 observations, 6,967 schools, 31 states) in all four years from
2022 through 2025.

To avoid compositional changes, especially between 2021 and the rest of
the panel, and in keeping with the analyses in
Section~\ref{sec-balancing}, most analyses in this paper use the
balanced panel of units present in each year from 2022-2025, but full
panel results are shown in Section~\ref{sec-unbalanced}. This limitation
does overrepresent schools and tests with more consistent testing
regimes in the data. The use of the balanced panel also somewhat changes
the distribution of exams across grade level, with only 26.1\% and
27.6\% of 9th and 12th grade exams being retained, while retention for
10th, 11th, and end-of-course exams ranged from 42.4\% to 53.6\%.

\FloatBarrier

\section{Does Google Trends track measured
traffic?}\label{sec-similarweb}

The exposure measure is built from a search index, not from usage.
Similarweb reports the quantity one would rather have, namely counts of
visits to chatgpt.com and openai.com, but only for 15 states and never
below the state level, which is why it cannot serve as the treatment
variable in a design that compares metropolitan areas. It can, however,
be used to check the source that can. The two series overlap on 2,340
state-weeks running from January 2023 through December 2025.

\begin{figure}

\centering{

\pandocbounded{\includegraphics[keepaspectratio]{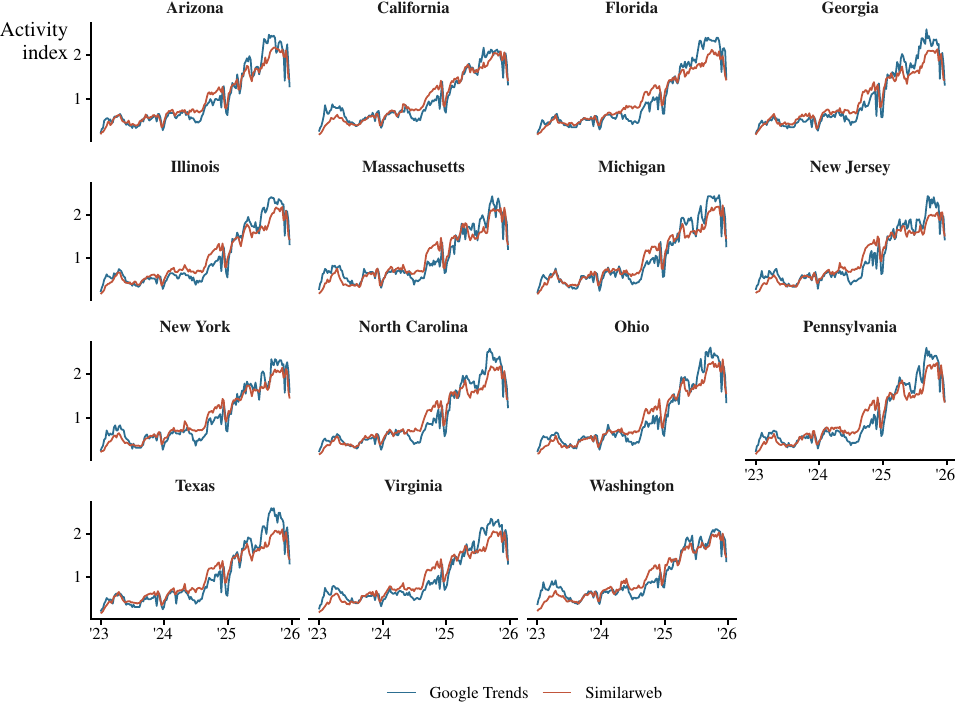}}

}

\caption{\label{fig-gtsw}ChatGPT activity from Google Trends and
Similarweb}

\end{figure}%

Figure~\ref{fig-gtsw} plots both series for each of the 15 states, with
each series scaled to a mean of one within its own state so that an
index and a visit count can be shown on the same axis. The two track
each other closely. Weekly correlations within state run from 0.937 to
0.968, with a median of 0.958. Part of this correlation is driven by a
general upward trend. Removing week fixed effects, the residual
correlation is 0.49, still substantial. Google Trends is therefore
carrying real cross-sectional information about relative activity.

\begin{figure}

\centering{

\pandocbounded{\includegraphics[keepaspectratio]{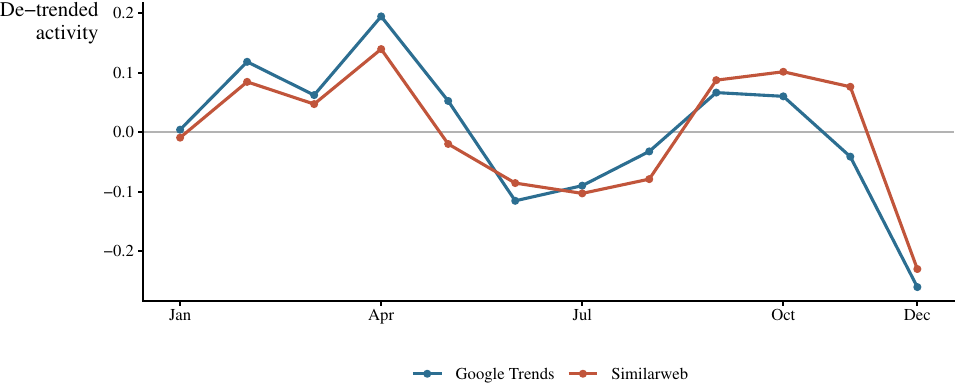}}

}

\caption{\label{fig-gtsw-seas}Seasonality in ChatGPT activity from
Google Trends and Similarweb}

\end{figure}%

Seasonality is what this paper actually uses, so agreement on the shape
of the year matters more than agreement on the level.
Figure~\ref{fig-gtsw-seas} averages each source by month after removing
a linear time trend separately within each state and source, which is
necessary because both series grow over the period and a common trend
would otherwise be read as a seasonal pattern. The two profiles are
close, correlating at 0.90 across the twelve months. Both show the
summer decline and both show the sharp December drop that coincides with
school holidays.

Summer-dip measures correlate strongly between data sources as well,
shown in Table~\ref{tbl-gtsw}. Pooling all 45 state-years, the two dips
correlate at 0.90. The within-year correlations across states are lower
but still substantial, at 0.84 in 2023, 0.55 in 2024 and 0.60 in 2025.
The Google Trends-based measure is capturing something real, since it
ranks states similarly to the web traffic-based measure.

\begin{table}

\caption{\label{tbl-gtsw}Agreement between Google Trends and Similarweb}

\centering{

\fontsize{7.0pt}{8.0pt}\selectfont
\begin{tabular*}{\linewidth}{@{\extracolsep{\fill}}lrrr}
\toprule
 & Pearson & Spearman & N \\ 
\midrule\addlinespace[2.5pt]
\multicolumn{4}{l}{A. Summer dip, by year} \\[2.5pt] 
\midrule\addlinespace[2.5pt]
2023 & 0.836 & 0.907 & 15 \\ 
2024 & 0.549 & 0.621 & 15 \\ 
2025 & 0.604 & 0.586 & 15 \\ 
All years pooled & 0.902 & 0.902 & 45 \\ 
\midrule\addlinespace[2.5pt]
\multicolumn{4}{l}{B. Underlying series} \\[2.5pt] 
\midrule\addlinespace[2.5pt]
Weekly activity, within state & 0.958 &  & 2,340 \\ 
Weekly activity, net of week effects & 0.485 &  & 2,340 \\ 
Seasonal profile, across months & 0.901 &  & 12 \\ 
\bottomrule
\end{tabular*}
\begin{minipage}{\linewidth}
Notes: Correlations between ChatGPT activity as measured by the Google Trends search index and by Similarweb visit counts to chatgpt.com and openai.com, across the 15 states for which Similarweb reports state-level traffic. Panel A applies the school and summer windows of Section 3.2 to each source separately and correlates the resulting ratios across states. Panel B reports the median within-state correlation of the weekly series, the same correlation after removing week fixed effects, and the correlation of the de-trended monthly seasonal profiles.\\
\end{minipage}

}

\end{table}%

\FloatBarrier

\section{Unbalanced panel results}\label{sec-unbalanced}

\FloatBarrier

The high school estimates in the body are computed on the panel balanced
over 2022 through 2025, which contains 22,117 units. This appendix
repeats each of them on the full unbalanced panel of 52,982 units, which
is the sample used in earlier drafts of this paper. Results on the
narrower panel balanced over 2021 through 2025 are available from the
author.

\begin{table}

\caption{\label{tbl-main-unbal}Effect of ChatGPT exposure on test
scores, unbalanced panel}

\centering{

\fontsize{7.0pt}{8.0pt}\selectfont
\begin{tabular*}{\linewidth}{@{\extracolsep{\fill}}lrr}
\toprule
Specification & Estimate & Obs \\ 
\midrule\addlinespace[2.5pt]
\multicolumn{3}{l}{A. High school (grades 9-12), unbalanced panel} \\[2.5pt] 
\midrule\addlinespace[2.5pt]
Baseline (all metros) & 0.0051 (0.0051) & 127,915 \\ 
Higher-precision metros only & 0.0038 (0.0079) & 93,117 \\ 
Weighted by test-takers & 0.0068 (0.0066) & 108,989 \\ 
Unwinsorised exposure & 0.0050 (0.0044) & 127,915 \\ 
Exposure purged of college share & 0.0046 (0.0050) & 127,915 \\ 
Local-calendar exposure & 0.0176 (0.0090)* & 127,915 \\ 
Proficiency-share outcome & 0.0012 (0.0021) & 127,915 \\ 
Metro covariates x year & 0.0026 (0.0059) & 126,631 \\ 
State + year FE (no state x year) & 0.0035 (0.0360) & 127,915 \\ 
Cluster on metro & 0.0051 (0.0055) & 127,915 \\ 
Drop 2021 (COVID testing year) & -0.0022 (0.0058) & 104,548 \\ 
Include 2025 & 0.0022 (0.0062) & 163,772 \\ 
\midrule\addlinespace[2.5pt]
\multicolumn{3}{l}{B. Grades 3-8 (age placebo)} \\[2.5pt] 
\midrule\addlinespace[2.5pt]
Mathematics & -0.0006 (0.0022) & 116,253 \\ 
Reading/LA & 0.0030 (0.0019) & 124,319 \\ 
Mathematics, + SES x year & -0.0001 (0.0022) & 111,495 \\ 
Reading/LA, + SES x year & 0.0041 (0.0020)* & 119,052 \\ 
\midrule\addlinespace[2.5pt]
\multicolumn{3}{l}{C. High school, time-varying exposure measure} \\[2.5pt] 
\midrule\addlinespace[2.5pt]
Baseline (all metros) & 0.0078 (0.0036)** & 140,344 \\ 
Higher-precision metros only & 0.0145 (0.0035)*** & 103,384 \\ 
Weighted by test-takers & 0.0050 (0.0028)* & 123,397 \\ 
Unwinsorised exposure & 0.0029 (0.0044) & 140,344 \\ 
Proficiency-share outcome & 0.0024 (0.0014)* & 140,344 \\ 
Metro covariates x year & 0.0012 (0.0038) & 138,950 \\ 
State + year FE (no state x year) & 0.0186 (0.0149) & 140,344 \\ 
Cluster on metro & 0.0078 (0.0030)*** & 140,344 \\ 
Drop 2025 (weak exposure measure) & -0.0007 (0.0047) & 104,487 \\ 
Fixed exposure, same years & -0.0065 (0.0081) & 140,405 \\ 
\midrule\addlinespace[2.5pt]
\multicolumn{3}{l}{D. High school, time-varying exposure by year} \\[2.5pt] 
\midrule\addlinespace[2.5pt]
2022 (reference) & 0 &  \\ 
2023 & 0.0112 (0.0049)** &  \\ 
2024 & 0.0061 (0.0032)* &  \\ 
2025 & 0.0283 (0.0190) &  \\ 
\bottomrule
\end{tabular*}
\begin{minipage}{\linewidth}
Notes: As in the corresponding table in the body, but on the full unbalanced panel over 2021-2024. Coefficients are test-score standard deviations per standard deviation of exposure, except the proficiency-share rows, which are shares on a 0-1 scale. Panel B, grades 3-8, is identical to the body version, since the SEDA panel is never restricted. Significance: * p<0.10, ** p<0.05, *** p<0.01.\\
\end{minipage}

}

\end{table}%

\begin{table}

\caption{\label{tbl-hetero-unbal}Effect heterogeneity, unbalanced panel}

\centering{

\fontsize{7.0pt}{8.0pt}\selectfont
\begin{tabular*}{\linewidth}{@{\extracolsep{\fill}}lrrrr}
\toprule
 & Estimate & p-value & States & Obs \\ 
\midrule\addlinespace[2.5pt]
\multicolumn{5}{l}{A. By subject and grade} \\[2.5pt] 
\midrule\addlinespace[2.5pt]
English, 9th Grade & -0.0066 (0.0168) & 0.704 & 10 & 4,709 \\ 
English, 10th Grade & 0.0168 (0.0093)* & 0.088 & 19 & 9,462 \\ 
English, 11th Grade & 0.0034 (0.0115) & 0.771 & 21 & 14,410 \\ 
English, End of course & 0.0219 (0.0162) & 0.206 & 11 & 14,742 \\ 
Mathematics, 9th Grade & 0.0008 (0.0172) & 0.962 & 10 & 4,445 \\ 
Mathematics, 10th Grade & 0.0092 (0.0096) & 0.352 & 15 & 4,901 \\ 
Mathematics, 11th Grade & 0.0065 (0.0174) & 0.714 & 23 & 13,353 \\ 
Mathematics, End of course & -0.0042 (0.0046) & 0.379 & 15 & 26,026 \\ 
Science, 9th Grade & 0.0411 (0.0010)*** & 0.000 & 5 & 1,108 \\ 
Science, 10th Grade & -0.0101 (0.0100) & 0.340 & 9 & 2,803 \\ 
Science, 11th Grade & -0.0176 (0.0230) & 0.455 & 20 & 9,531 \\ 
Science, 12th Grade & 0.0663 (0.0000)*** & 0.000 & 5 & 4,185 \\ 
Science, End of course & -0.0074 (0.0066) & 0.280 & 14 & 17,801 \\ 
\midrule\addlinespace[2.5pt]
\multicolumn{5}{l}{B. Schools, by baseline performance tercile} \\[2.5pt] 
\midrule\addlinespace[2.5pt]
Bottom third & 0.0166 (0.0102) & 0.113 &  & 37,185 \\ 
Middle third & 0.0314 (0.0151)** & 0.044 &  & 37,411 \\ 
Top third & -0.0021 (0.0130) & 0.872 &  & 37,024 \\ 
Bottom minus top & 0.0187 (0.0199) & 0.353 &  &  \\ 
\midrule\addlinespace[2.5pt]
\multicolumn{5}{l}{C. Students, within-school distribution} \\[2.5pt] 
\midrule\addlinespace[2.5pt]
Share above lowest band & 0.0011 (0.0036) & 0.759 &  & 117,820 \\ 
Share in highest band & -0.0013 (0.0010) & 0.197 &  & 117,816 \\ 
\bottomrule
\end{tabular*}
\begin{minipage}{\linewidth}
Notes: As in the corresponding table in the body, but on the full unbalanced panel. States is the number of clustering states behind each row of Panel A. A comparison of the top and bottom tercile coefficients using randomization inference gives p = 0.398 . Panel C is restricted to tests reporting at least four proficiency bands, with outcomes as shares on a 0-1 scale. Significance: * p<0.10, ** p<0.05, *** p<0.01.\\
\end{minipage}

}

\end{table}%

\begin{table}

\caption{\label{tbl-bounds-unbal}Bounds on the effect of ChatGPT
exposure, unbalanced panel}

\centering{

\fontsize{7.0pt}{8.0pt}\selectfont
\begin{tabular*}{\linewidth}{@{\extracolsep{\fill}}lrlrr}
\toprule
Sample & Estimate & 95\% CI & Bound & Bound, corrected \\ 
\midrule\addlinespace[2.5pt]
High school, all subjects & 0.0051 & [-0.0052, 0.0154] & 0.0154 & 0.0193 \\ 
High school, higher-precision metros & 0.0038 & [-0.0122, 0.0197] & 0.0197 & 0.0214 \\ 
Grades 3-8, mathematics & -0.0006 & [-0.0051, 0.0039] & 0.0051 & 0.0064 \\ 
Grades 3-8, reading & 0.0030 & [-0.0009, 0.0069] & 0.0069 & 0.0086 \\ 
\bottomrule
\end{tabular*}
\begin{minipage}{\linewidth}
Notes: Estimates and intervals are in test-score standard deviations per standard deviation of exposure. The bound column is the larger absolute end of the 95\% confidence interval. The corrected column divides by the classical attenuation factor for classical measurement error in a regressor. The full panel is used.\\
\end{minipage}

}

\end{table}%

\FloatBarrier

\section{Impacts on state-level college entrance examination
scores}\label{sec-satfail}

State-level SAT data is available from 2017 onwards, so a placebo is
testable over several years. The placebo test for SAT scores fails.
High-exposure states sit roughly 0.10 standard deviations above their
2022 level in each of 2017, 2018 and 2019, declining monotonically
toward 2022. A joint test that the pre-2022 coefficients are zero
rejects at \(p = 0.021\). States with high levels of educational ChatGPT
usage already showed different SAT score trends before ChatGPT existed.

This test uses only SAT scores since the available participation rate
data does not go this far back. However, this test treats equally states
in which the SAT is very common and states in which most students take
the ACT. Restricting the sample to the 27 states in which the SAT is the
more popular of the two exams, the placebo test still fails. The joint
test of the pre-2022 coefficients rejects at \(p = 0.045\).

Given these violations, and the overall small sample size and thus low
power of the test, any result for these college entrance exam results
would not be believable. However, for completeness, post-treatment
estimates are shown in Table~\ref{tbl-satblend}, using both SAT-only and
participation-weighted SAT/ACT blended scores. All results are
statistically insignificant.

\begin{table}

\caption{\label{tbl-satblend}SAT-only and blended SAT+ACT outcomes,
identical samples}

\centering{

\fontsize{7.0pt}{8.0pt}\selectfont
\begin{tabular*}{\linewidth}{@{\extracolsep{\fill}}lrrrr}
\toprule
Exam & Years & SAT only & Blended SAT+ACT & Obs \\ 
\midrule\addlinespace[2.5pt]
Total & 2021-2024 & -0.002 (0.048) & -0.020 (0.033) & 204 \\ 
Math & 2022-2024 & 0.018 (0.044) & 0.000 (0.028) & 153 \\ 
ERW & 2022-2024 & 0.034 (0.053) & -0.011 (0.034) & 153 \\ 
\bottomrule
\end{tabular*}
\begin{minipage}{\linewidth}
Notes: The blended outcome is the participation-weighted average of the state SAT mean and the state ACT mean converted to the SAT scale by the official concordance, standardized with the SAT-only mean and standard deviation of the same exam so that the two columns are in the same units. The years differ across rows because ACT reports the composite from 2021 but subject scores only from 2022. These results should not be read as causal effects. Standard errors clustered on state in parentheses. Significance: * p<0.10, ** p<0.05, *** p<0.01.\\
\end{minipage}

}

\end{table}%

\end{document}